\documentclass{aa}
\usepackage[varg]{txfonts}
\usepackage{subcaption}
\usepackage{gensymb}
\usepackage{epstopdf}
\usepackage{graphicx}
\usepackage{gensymb}
\usepackage{subcaption}
\usepackage{wrapfig}
\usepackage{nicefrac}
\graphicspath{{./Figures/}}
\makeatletter
\newcommand{\RNum}[1]{\uppercase\expandafter{\romannumeral #1\relax}}

\begin{document}

\title{Sunspot rotation}
\subtitle{\RNum{2}. Effects of varying the field strength and twist of an emerging flux tube}
\author{Z. Sturrock
  \and A. W. Hood}
  
\institute{School of Mathematics and Statistics, University of St Andrews, St Andrews, Fife, KY16 9SS, UK\\
email: \texttt{zoe.sturrock@st-andrews.ac.uk} } 
\date{Received 22 February 2016 / Accepted 12 May 2016}

\abstract{Observations of flux emergence indicate that rotational velocities may develop within sunspots. However, the dependence of this rotation on sub-photospheric field strength and twist remains largely unknown.}{We investigate the effects of varying the initial field strength and twist of an emerging sub-photospheric magnetic flux tube on the rotation of the sunspots at the photosphere.}{We consider a simple model of a stratified domain with a sub-photospheric interior layer and three overlying atmospheric layers. A twisted arched flux tube is inserted in the interior and is allowed to rise into the atmosphere. To achieve this, the MHD equations are solved using the Lagrangian-remap code, \emph{Lare3d}. We perform a parameter study by independently varying the sub-photospheric magnetic field strength and twist.}{Altering the initial magnetic field strength and twist of the flux tube significantly affects the tube's evolution and the rotational motions that develop at the photosphere. The rotation angle, vorticity, and current show a direct dependence on the initial field strength. We find that an increase in field strength increases the angle through which the fieldlines rotate, the length of the fieldlines extending into the atmosphere, and the magnetic energy transported to the atmosphere. This also affects the amount of residual twist in the interior. The length of the fieldlines is crucial as we predict the twist per unit length equilibrates to a lower value on longer fieldlines. No such direct dependence is found when we modify the twist of the magnetic field owing to the complex effect this has on the tension force acting on the tube. However, there is still a clear ordering in quantities such as the rotation angle, helicity, and free energy with higher initial twist cases being related to sunspots that rotate more rapidly, transporting more helicity and magnetic energy to the atmosphere.}{}

\maketitle

\section{Introduction}

 Sunspots are large concentrations of magnetic flux that appear on the solar surface and are due to the emergence of flux from the convection zone. Rotational motions regularly develop within sunspots in both observations and numerical simulations. This phenomena can trigger explosive events in the Sun's atmosphere such as coronal mass ejections (CMEs) and solar flares. There has been much debate over the driver of this rotation; several theories have been introduced including photospheric flows caused by differential rotation and magneto-convective dynamics~\citep{Brown2003} and the flux emergence process \citep{Min2009,Brown2003}. 
 
 Observations of sunspot rotation date back to the beginning of the twentieth century when~\cite{Evershed1909} first found spectral observations that exhibited signs of rotating sunspots. Since their discovery, several observational studies have been conducted \citep{Brown2003,Yan2007,Yan2008,Min2009}. Through the tracking of prominent features in magnetograms, these studies have shown that sunspots can undergo rotations of the order of a few hundreds degrees over a period of a few days. We direct the reader to~\cite{Sturrock2015} (henceforth referred to as Paper I) for a more detailed review of recent observations of sunspot rotation.

Following on from the first paper in the series, Paper I, we continue to investigate the process of flux emergence as a cause for sunspot rotation. The net torque produced by magnetic flux emergence was found to be the mechanism for the rotational movement of sunspots. In this paper, we focus on how a change in the parameters determining the magnetic structure of the flux tube affect the evolution of flux and rotation of the spots. Similar flux emergence parameter studies have been conducted in the past with a slightly different focus.~\cite{Murray2006} investigated the effects of varying the magnetic field strength and twist of a sub-photospheric cylindrical flux tube with the aim of understanding how these parameters affect the evolution of the flux tube on its rise to the atmosphere. The authors found a self-similar evolution in the rise and emergence of the tube when the magnetic field strength is varied. However, they did not find such a simple self-similar evolution when varying the twist due to the non-linear interaction between the twist and the tension force acting on the tube. Nevertheless, if the field strength or twist is low enough, the flux tube cannot fully emerge into the atmosphere. Another interesting parameter study was carried out by~\cite{MacTaggart2009a} where they used a different initial condition, namely a toroidal flux tube. The advantages of modelling a flux tube in the shape of a half-torus, and hence a tube rooted deeper in the interior are two-fold: the axis is able to fully emerge in cases prohibited by the cylindrical model and the sunspot pair does not continue to drift apart. We use this model as our initial condition and instead focus on how varying the above parameters affects the features of sunspot rotation.

Although sunspot rotation has been a very attractive topic to both observers and theorists in recent years, the rotation's dependence on field strength and twist of the initial sub-photospheric field has, to the best of our knowledge, been left unexplored. Given that there are no current observations of sunspot rotation for varying initial strengths and twist, the results we find based on this simple model should be checked against future observations. We can, however, make predictions on the effect of these parameters from previous studies~\citep{Murray2006,MacTaggart2009a}. For instance, we predict that the twist of the sub-photospheric tube should have a substantial effect on the rate at which the sunspots rotate as we expect a larger rotation for more highly twisted fields, as there is more twist stored in the interior field to unwind. However, as we will discuss later, the density deficit's non-linear dependence on the twist, $\alpha$, will complicate this effect. In addition, we predict that the role of the magnetic field strength may not be so important to the magnetic flux tube's rotation at the photosphere. Nonetheless, the density deficit is directly proportional to the field strength of the tube squared, and hence the stronger fields emerge more fully in our experiments. This allows the axis to align vertically which may impact on the rotation. By considering the effects of changing the field strength we actually study the effects of a differing emerging rate.

In this paper, we present results from $3$D MHD simulations of buoyant magnetic flux tubes rising through the solar interior and emerging at the photosphere. We are particularly interested in the rotational motions of the photospheric footpoints of the tube, i.e. the sunspots. We have varied two of the parameters defining the magnetic structure of the sub-photospheric flux tube, namely the magnetic field strength at the axis, $B_0$, and the twist of the tube, $\alpha$. Our aim is to identify the effect of these parameters on a number of quantities relating to the rotation of sunspots. In addition, we seek to understand the process controlling the amount of sunspot rotation. To determine the individual effect of these parameters, we vary the field strength and twist, independently of each other.
 
The remainder of this paper is laid out as follows. In Section~\ref{sec:modelsetup}, we briefly outline the model setup and introduce the parameters we choose for the parametric study. A short outline of the general dynamics and main phases of sunspot rotation from a general experiment is given in Section~\ref{sec:general}. Section~\ref{sec:varyB0} and Section~\ref{sec:varyalpha} present the results of varying the magnetic field strength and twist of the flux tube respectively. Finally, we conclude the paper with a summary of the results and main conclusions in Section~\ref{sec:conclude}.
 
\section{Model setup}
\label{sec:modelsetup}
For the experiments performed in this paper we use the same model initial set-up as defined in Paper I where we solve the 3D resistive MHD equations using the \emph{Lare3d} code~\citep{Arber2001}. Precisely, we model a stratified background atmosphere consisting of a sub-photospheric layer, an isothermal photosphere, a transition region, and an isothermal lower coronal layer. We then insert a twisted toroidal flux tube, introduced by~\cite{Hood2012}, given in Cartesian coordinates as
\begin{eqnarray}
B_x  & = & - B_{\theta}(r)\frac{R-R_0}{r},\nonumber\\
B_y & = & - B_{\phi}(r)\frac{z-z_{\text{base}}}{R} + B_{\theta}(r)\frac{x}{r}\frac{y}{R},\nonumber\\
B_z & = & B_{\phi}(r) \frac{y}{R} + B_{\theta}(r)\frac{x}{r}\frac{z-z_{\text{base}}}{R},
\label{eq:initialb}
\end{eqnarray}
where $B_{\phi}=B_0e^{-r^2/a^2}$ and $B_{\theta}=\alpha rB_{\phi}$. $B_0$ is the axial field strength, $\alpha$ is the degree of twist of the fieldlines, $a$ is the minor radius of the flux tube, and $R_0$ is the major radius. We define the axis of the flux tube to be the fieldline threading through the centre at $r=0$.

We must also highlight the difference in pressure and density of the flux tube from its surroundings as this has important consequences later. In order to reach an equilibrium we must establish a pressure excess, $p_{\text{exc}}$. We make the flux tube buoyant by introducing a density excess by maintaining the temperature and pressure excess as given by
\begin{equation}
\rho_{\text{exc}}(r) = \frac{p_{\text{exc}}}{T(z)} = \frac{B_0^2}{4T(z)}e^{-2r^2/a^2}(\alpha^2a^2 - 2\alpha^2r^2 -2).
\label{eq:densitydef}
\end{equation}
We direct the reader to Paper I for further details. We note that the pressure excess, and in turn density excess, is negative for all $r$ if $\alpha a < \sqrt{2}$. In all of our parameter choices this is satisfied and so the density excess is negative and is instead a density deficit. Equivalently, this means that the outwardly directed magnetic pressure force is larger than the inwardly directed magnetic tension force, and hence the gas pressure gradient acts inwardly to balance the forces.

\begin{figure}[ht]
\centering
\includegraphics[scale=0.27]{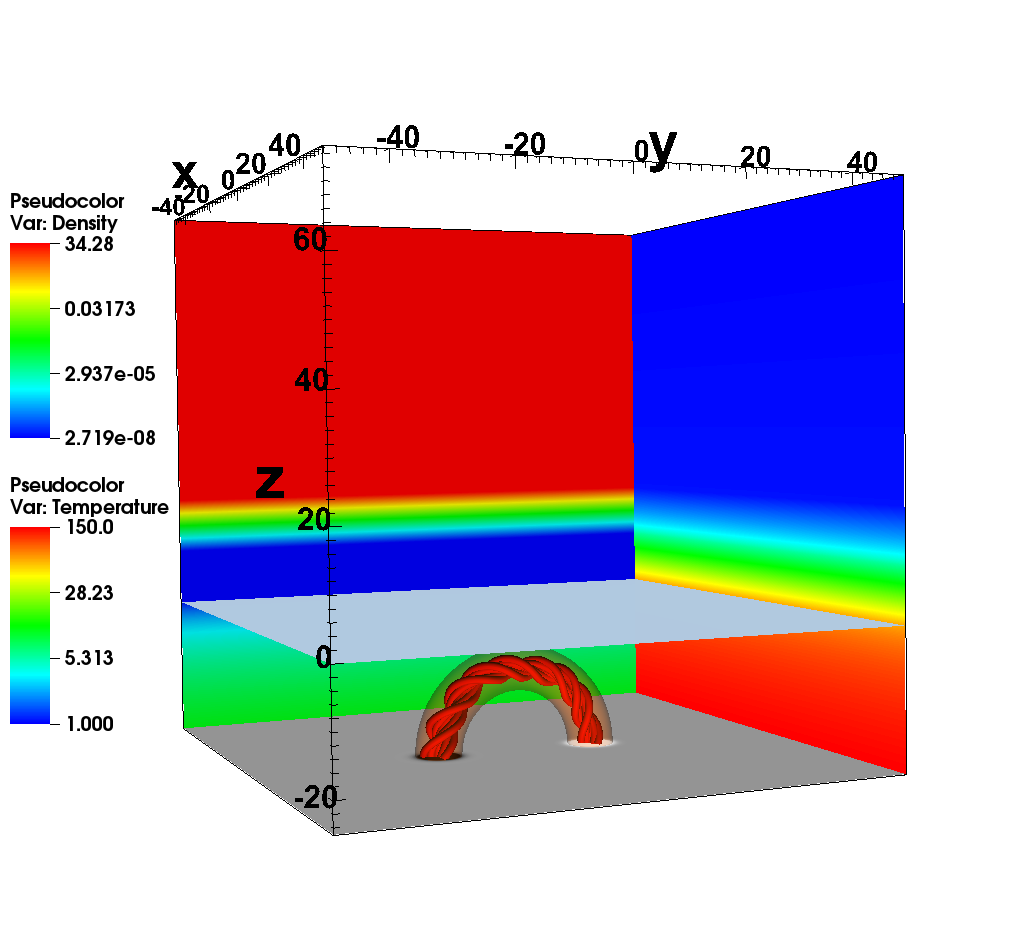}
\caption{Summary of initial setup. The background density distribution is shown on the right wall, the temperature distribution on the back wall, a selection of fieldlines are shown in red, and an isosurface of the magnetic field ($|\mathbf{B}|=1$) is over-plotted. The $z=0$ plane is also highlighted in grey as the solar surface.}
\label{fig:rotinitialsetup}
\end{figure}

\subsection{Parameter choice}
\label{sec:parameters}

In this study, we fix the base of the computational domain at $z_{\text{base}}=-25$. The major radius of the torus is $R_0=15$ and the minor radius is $a=2.5$ for all experiments. We choose to vary the magnetic field strength at the axis of the interior tube, $B_0$, and the degree of twist, $\alpha$. The twist is assumed to be positive and constant in all experiments, ensuring that all tubes are uniformly right-hand twisted. Each fieldline rotates about the axis through an angle of $\alpha$ radians over one unit of distance along the axis. The initial set-up of a representative experiment is summarised in Figure~\ref{fig:rotinitialsetup}, with parameters $B_0=7$ (axial field strength of $9100$~G) and $\alpha = 0.4$ (three full turns of twist in interior tube).

The experiments are split into two groups: Group 1 where $\alpha$ is kept fixed and $B_0$ is varied and Group 2 where $B_0$ is fixed and $\alpha$ is varied. A summary of the $B_0$ and $\alpha$ values under consideration is given in Table~\ref{table:parameters}. It should be noted we only consider a relatively small range of $B_0$ values as a consequence of our model choice. As found in previous parameter studies (see~\citealp{MacTaggart2009a}), if we pick a lower $B_0$ value, the flux tube will fail to fully emerge, and if we choose a much higher $B_0$ value, we may encounter unphysical negative pressures. In fact, if we increase the initial axial field strength to $B_0=12$, we encounter negative pressure in the particular initial set-up outlined in the previous section. It is, however, important to note that the range of allowable field strengths will vary from set-up to set-up. The twist values range from $\alpha=0.2$ corresponding to a turn and a half of twist to $\alpha=0.4$ that corresponds to three full turns of twist in the initial field. The total flux threading a cross section of the tubes we study range from $3.7\times10^{19}$~Mx to $3.7\times10^{20}$~Mx. This is typical of a small active region or large ephemeral region. 
\begin{table}[ht]
\centering
\caption{Parameter space under investigation.}
\label{table:parameters}
\begin{tabular}{cc}
\hline\hline
\hspace{1cm} Group 1 \hspace{1cm} & \hspace{1cm} Group 2 \hspace{1cm} \\ [0.5ex]
 \hline\hline
 $B_0=[5,6,7,8,9,10]$ & $B_0=7$ \\[0.5ex]
 $\alpha=0.3$ & $\alpha=[0,2,0.3,0.4]$\\ 
 \hline
\end{tabular}
\end{table}

All experiments have been performed for different numbers of normalised time units, and hence different final times. The Group $1$ experiments are carried out as follows: $B_0=5$ for $216$ time units ($90$ minutes); $B_0=6$ for $180$ time units ($75$ minutes); $B_0=7$ for $154$ time units ($64$ minutes); $B_0=8$ for $135$ time units ($56$ minutes); $B_0=9$ for $120$ time units ($50$ minutes); and $B_0=10$ for $108$ time units ($45$ minutes). This ensures all experiments are performed for the same rescaled time, i.e. $\bar{t}=t\times B_0$. More details of rescaling the time follow in Section~\ref{sec:varyB0}. The Group $2$ experiments, on the other hand, are all executed for the same number of time units, $120$ time units or equivalently $50$ minutes.

\begin{figure}[!ht]
\centering
\begin{subfigure}{0.48\textwidth}
\centering
\includegraphics[scale=0.4]{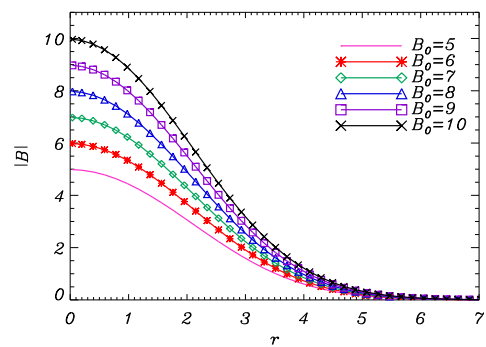}
\subcaption{Group $1$}
\label{subfig:bmagb0}
\end{subfigure}\\
\begin{subfigure}{0.48\textwidth}
\centering
\includegraphics[scale=0.4]{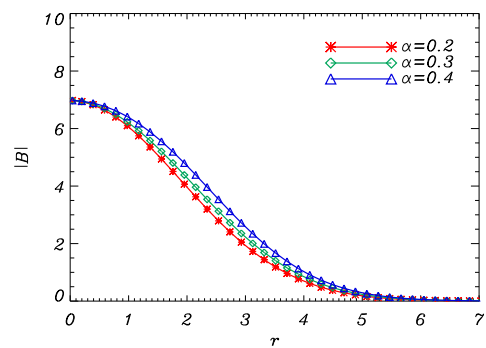}
\subcaption{Group $2$}
\label{subfig:bmagalpha}
\end{subfigure}
\caption{Radial distribution of the initial magnetic field strength, $\mathbf{|B|}$, at $z=-25$ for varying (a) $B_0$ and (b) $\alpha$.}
\end{figure}

\begin{figure}[ht]
\centering
\begin{subfigure}{0.48\textwidth}
\centering
\includegraphics[scale=0.4]{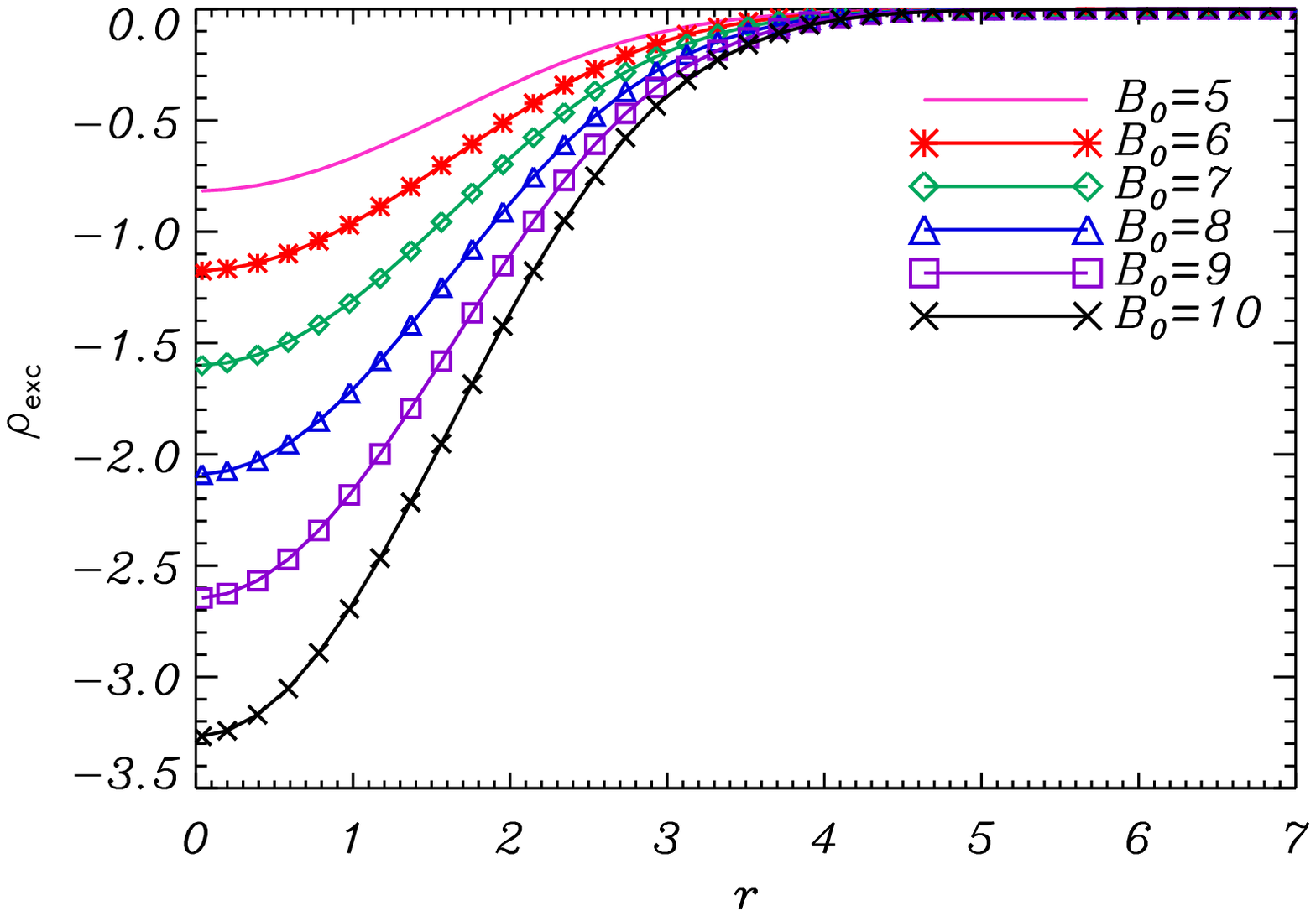}
\subcaption{Group $1$}
\label{subfig:rhodefb0}
\end{subfigure}\\
\begin{subfigure}{0.48\textwidth}
\centering
\includegraphics[scale=0.4]{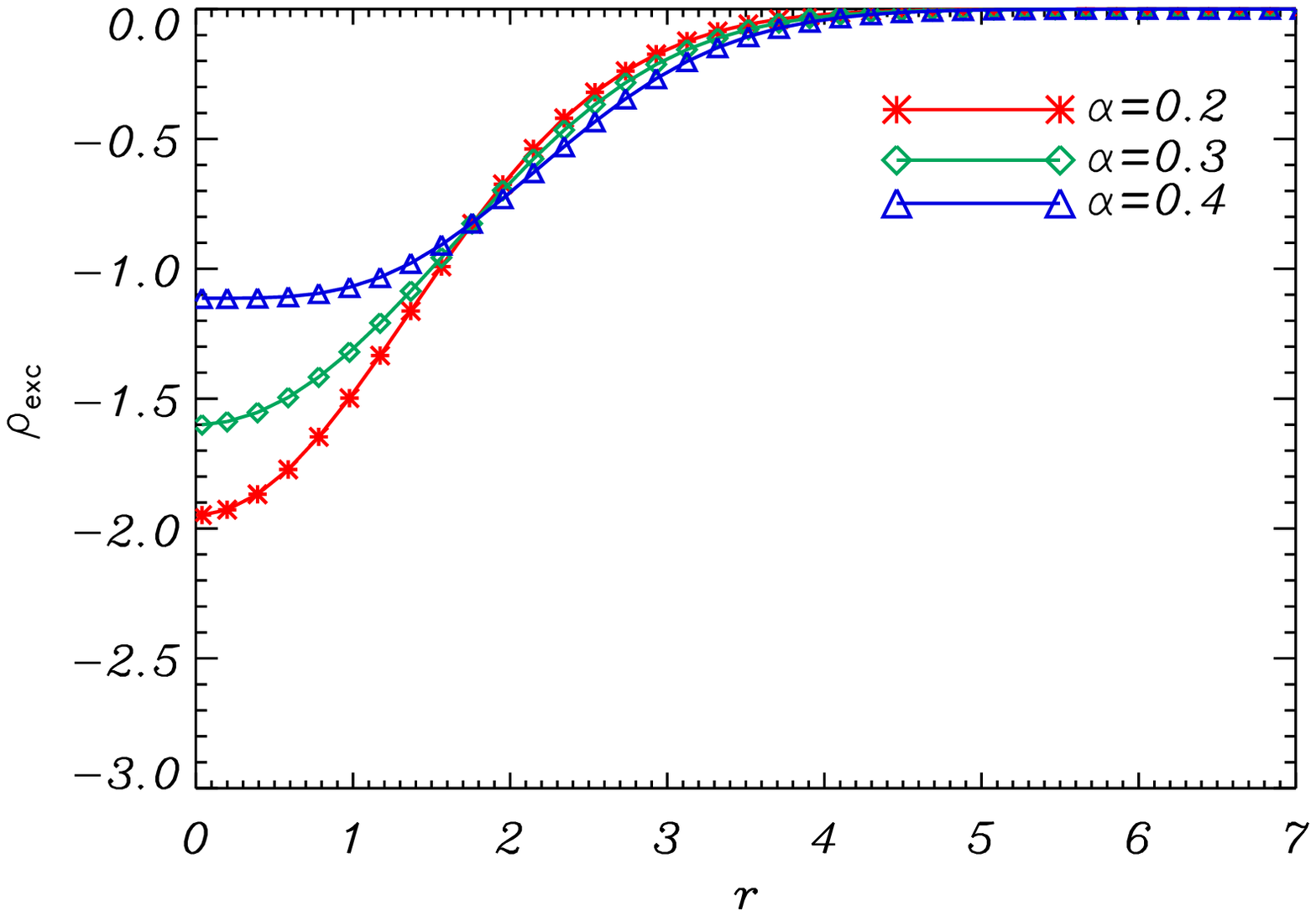}
\subcaption{Group $2$}
\label{subfig:rhodefalpha}
\end{subfigure}
\caption{Radial distribution of the initial density excess, $\rho_{\text{exc}}$, at $z=-25$ for varying (a) $B_0$ and (b) $\alpha$.}
\end{figure}

Before we proceed to discuss the results of our experiments, we analyse how the parameter choices affect the structure of the magnetic field and resulting density. From the density excess in Equation~\eqref{eq:densitydef}, it is clear that both the twist and magnetic field strength play an important role in controlling the buoyancy of the flux tube. For Group 1, the variation in $B_0$ significantly affects the magnetic field strength for all radii until a significant distance from the axis is reached, i.e. the edge of the tube, as shown in Figure~\ref{subfig:bmagb0}. The initial field strength of the tube is, of course, directly proportional to $B_0$. The buoyancy profile, too, is strongly dependent on $B_0$ because the density deficit is proportional to $B_0^2$. The $B_0=10$ tube will therefore be $4$ times more buoyant than the $B_0=5$ tube (see Figure~\ref{subfig:rhodefb0}).

In Group 2, the variation in $\alpha$ has a small effect on the field strength for all radii, leaving the field strength at the axis of the tube unchanged, as displayed in Figure~\ref{subfig:bmagalpha}. However, the more strongly twisted case has a slightly larger field strength as we move away from the axis. The buoyancy profile, on the other hand, is influenced by the value of $\alpha$ as increasing the amount of twist increases the inward acting magnetic tension force more than the outward acting magnetic pressure force, therefore altering the Lorentz force and in turn the density deficit. This results in the higher $\alpha$ cases having a smaller inwardly acting gas pressure gradient, and hence being less buoyant at the centre of the flux tube, as demonstrated in Figure~\ref{subfig:rhodefalpha}. At the outside of the flux tube the larger field strength of the higher $\alpha$ case causes the tubes to be more buoyant.

\section{General Dynamics}
\label{sec:general}

In this section, we discuss the general dynamics and features of an experiment with axial field strength $B_0=9$ and twist $\alpha=0.4$. For a full analysis of the properties of sunspot rotation in this particular experiment, we refer the reader to Paper I. The flux tube begins to rise buoyantly when the experiment begins due to the density deficit introduced. The decreasing temperature profile in the solar interior permits the flux tube to continue to rise until it reaches the photosphere. Due to the photosphere's stable stratification, the flux tube is no longer buoyant past this point. In order for the flux tube to fully emerge, the initiation of a secondary instability is necessary. If the field is strong enough, as in the case of this experiment, the flux tube then rises into the corona by means of the magnetic buoyancy instability. Further details of this instability can be found in the recent flux emergence review paper,~\cite{Hood2012}.

\begin{figure}[ht]
\centering
\begin{subfigure}{0.48\textwidth}
\centering
\includegraphics[scale=0.35]{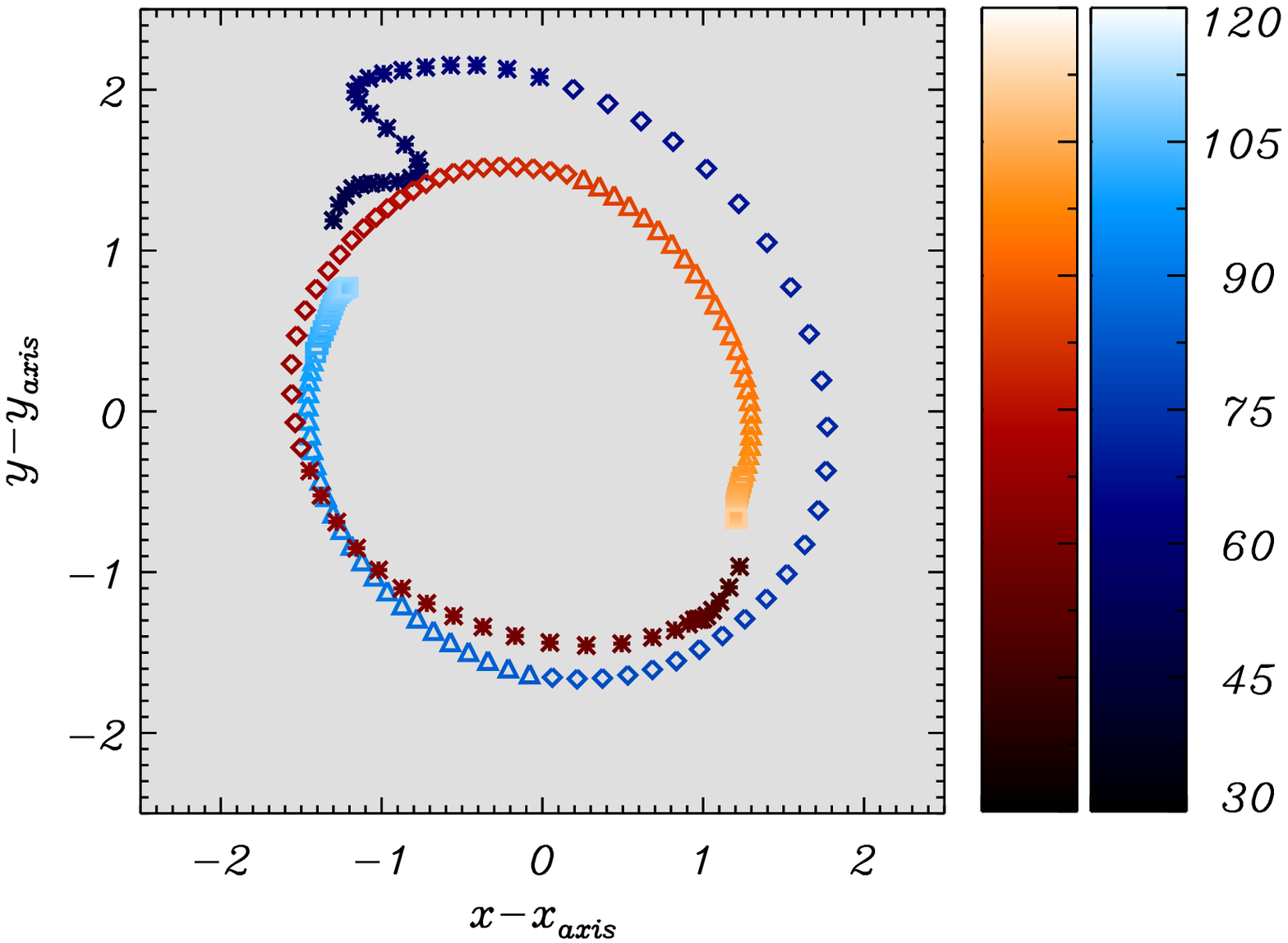}
\subcaption{}
\label{subfig:generaltraj}
\end{subfigure}\\
\begin{subfigure}{0.48\textwidth}
\centering
\includegraphics[scale=0.35]{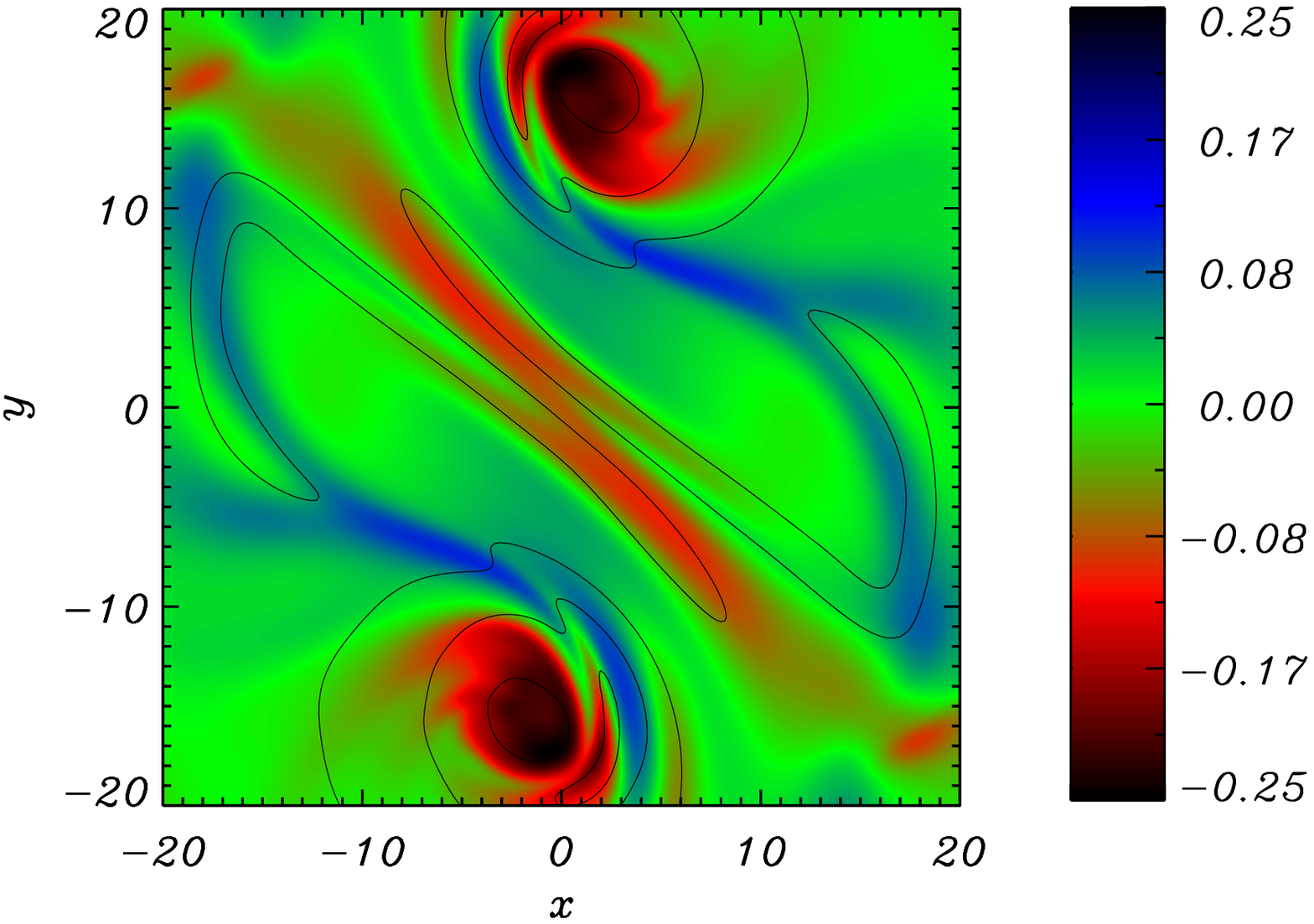}
\subcaption{}
\label{subfig:generalvortz}
\end{subfigure}
\caption{Analysis of sunspot rotation based on an experiment from Paper I~\citep{Sturrock2015}. (a) The trajectories of two fieldlines relative to the axis (centre of sunspot) with time coloured in the key. (b) Coloured contour of vertical vorticity, $\omega_z$, at a time soon after the intersecting field becomes vertical.}
\end{figure}

Once the field emerges and expands into the low density atmosphere, both the magnetic field and plasma exhibit signs of rotation within the two photospheric polarity sources. In this particular experiment, both sunspots undergo rotations through angles of up to $353\degree$. Particular fieldlines are traced in order to follow the trajectories through the photospheric plane as shown in Figure~\ref{subfig:generaltraj}. Following the approach introduced in Paper I, this figure shows the position of the fieldlines relative to the central axial fieldline and clearly shows the fieldlines follow an almost circular path as they rotate around the sunspot centre. This is also evidenced by vortical motions that develop within both sunspot centres as displayed in Figure~\ref{subfig:generalvortz}, indicating that the plasma within the sunspots are rotating clockwise. This has important ramifications for both the interior and atmospheric field. We find a clear depletion in magnetic helicity and energy in the interior portion of the volume as the rotational photospheric motions untwist the interior field. In the atmosphere, on the other hand, there is a steady increase in magnetic helicity and energy owing to both the emergence of magnetic flux and rotational motions at the photosphere. In total, $3.6\times10^{23}$~Wb$^2$ of helicity and $8.2\times10^{22}$~J of energy are transported to the atmosphere. We believe that this is related to the propagation of a torsional Alfv\'{e}n wave at the instance of emergence that untwists the twisted interior flux tube in an effort to equilibrate the twisted interior with the stretched corona~\citep{Longcope2000}. Now that the basic model and general behaviour of sunspot rotation is established, we consider two parameter groups introduced in Table~\ref{table:parameters}.

\section{Varying $B_0$ with fixed $\alpha$}
\label{sec:varyB0}
In this group, we fix $\alpha$ at $0.3$ ($\displaystyle{2\nicefrac{1}{4}}$ turns of twist or equivalently one turn in $567~\text{km}$) and vary the field strength at the axis of the tube, $B_0$, from $B_0=5$ ($6500$~G) to $B_0=10$ ($13000$~G) in steps of $1$ ($1300$G). This allows us to understand the sole effect that the magnetic field strength has on the rotation of the sunspots. Altering the initial field strength, $B_0$, changes the evolution of the field in two ways. Firstly, changing $B_0$ alters the initial density deficit thereby controlling the speed at which the flux tube rises through the solar interior (see Figure~\ref{subfig:rhodefb0}). Secondly, the tube's evolution is altered on its journey from the photosphere. In order for the flux tube to emerge, the magnetic buoyancy instability must be triggered which occurs when the plasma $\beta$ is lowered to one. This occurs more quickly for stronger fields due to their higher magnetic pressure. For a weaker field, on the other hand, the magnetic pressure is built up more slowly as the flux tube is squashed and the field is spread at the photosphere. Rotational motions are manifested in several different ways, and hence we investigate a variety of different quantities. Before delving into the rotational properties, we first analyse the general evolution of the magnetic field.

\subsection{General evolution}
\begin{figure}[ht]
\centering
\begin{subfigure}{0.48\textwidth}
\centering
\includegraphics[scale=0.4]{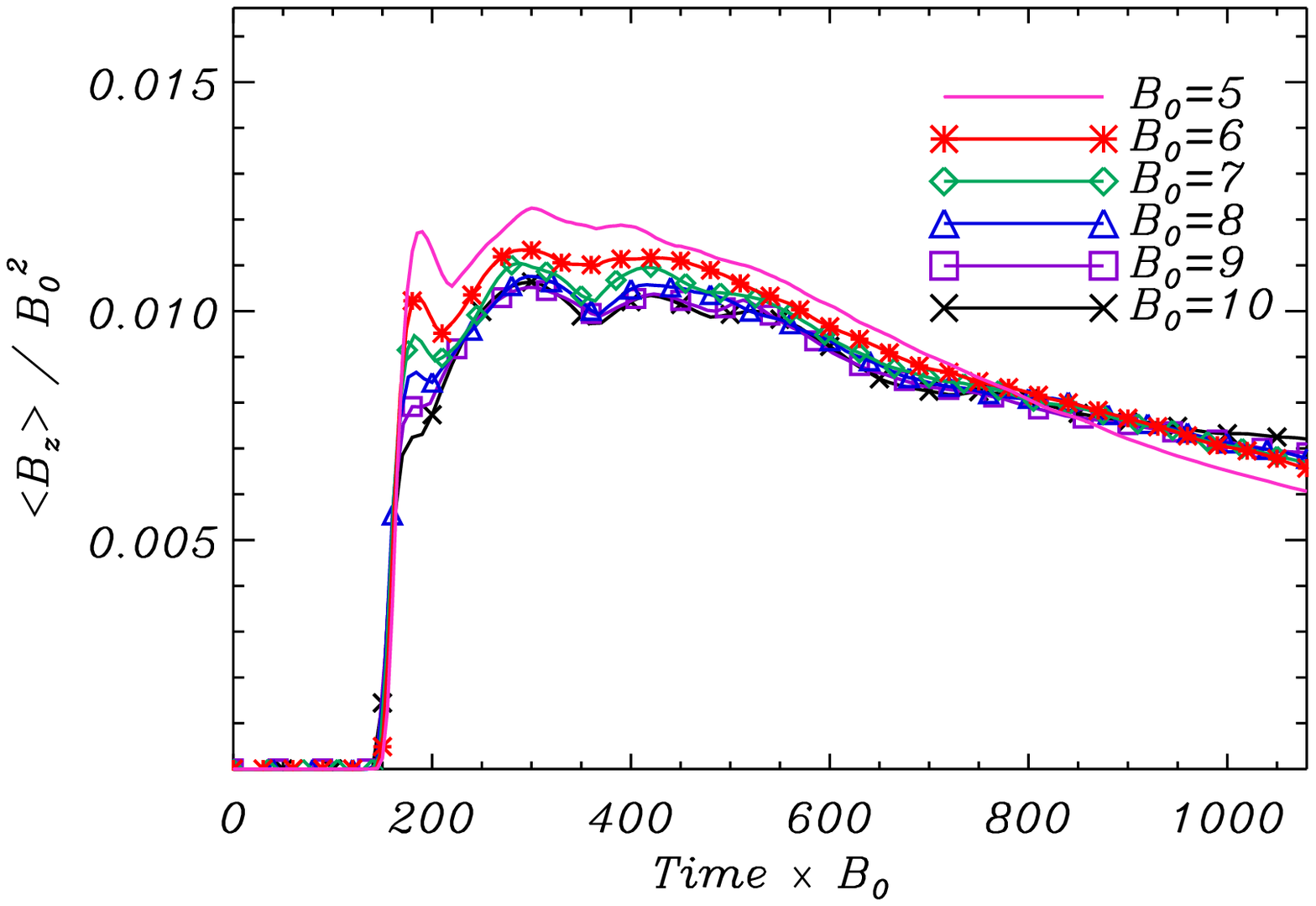}
\subcaption{}
\label{subfig:meanbzb0}
\end{subfigure}\\
\begin{subfigure}{0.48\textwidth}
\centering
\includegraphics[scale=0.4]{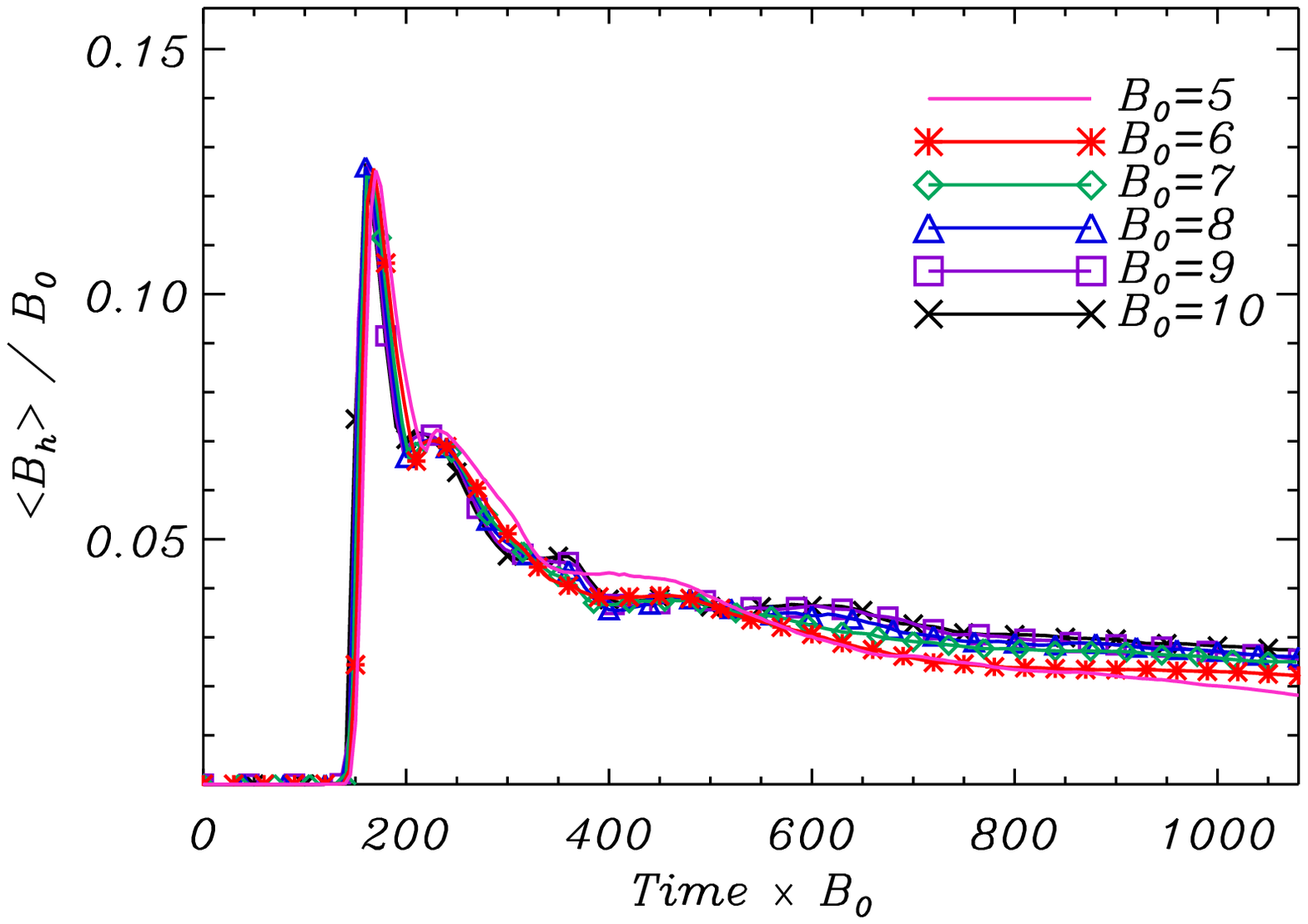}
\subcaption{}
\label{subfig:meanbhb0}
\end{subfigure}
\label{fig:generalrot1}
\caption{The evolution of (a) the scaled mean of $B_z$ on the $z=0$ plane and (b) the scaled mean of $B_h$ on the $z=0$ plane over rescaled time for the parameters outlined in the legend for Group 1.}
\end{figure}

\noindent To try and understand the influence of the interior magnetic field strength on the evolution of the tube as it rises, we consider how this affects the magnetic field strength at the photosphere. We consider two proxies for the magnetic field strength at the photosphere, namely the mean vertical field strength, ${\langle{B_z}\rangle}_{z=0}$, and the mean horizontal field strength, ${\langle{B_h}\rangle}_{z=0}$. Both expressions have been plotted in Figures~\ref{subfig:meanbzb0} and~\ref{subfig:meanbhb0} over rescaled time for all six Group $1$ simulations as coloured by the key. We calculate the mean by averaging over the photospheric region where $B_z > 3/4\text{max}(B_z)$. Hence, we assume that contributions from the main sunspot field are included and weaker undular field regions outside the spots are excluded. This has been compared with other proxies for the magnetic field, such as the maximum field strength and we find the same general behaviour in this case. In order to take into account the density deficit's dependence on $B_0$, we rescale the horizontal axis by redefining time as $\bar{t}=B_0t$. This is equivalent to measuring time on an Alfv\'{e}n timescale rather than a sound timescale. Before we proceed, we note that the symbols on this plot and all subsequent plots do not reflect the spatial or temporal resolution of the experiments. Unless stated otherwise, symbols are only plotted every five grid points (or time units) for visualisation purposes.

All three components of the initial magnetic field are proportional to $B_0$ and as such we may expect that the field will still be proportional to $B_0$ when the tube reaches the photosphere. Interestingly, from Figure~\ref{subfig:meanbzb0}, we find that the vertical field strength at the photosphere can instead be scaled by $B_0^{2}$, suggesting that stronger initial fields tend to concentrate and strengthen in the vertical direction at the photosphere. We find that stronger fields emerge more fully with a vertical axis, and hence possess a larger vertical field, $B_z$, at the photosphere. Flux tubes with weaker fields, on the other hand, tend to spread at the photosphere before the magnetic buoyancy instability is initiated. This could be responsible for a smaller than expected $B_z$ at the photosphere for lower $B_0$ values. 

For completeness, we have also included the horizontal field strength, $B_h=\sqrt{B_x^2+B_y^2}$ and find that it is proportional to $B_0$, as we predict. As we are averaging over the region where $B_z>3/4\text{max}(B_z)$, we do not see the effects of the horizontal expansion of the field for weaker $B_0$ values. It is important that we bear these scalings in mind when analysing later results as the magnetic field is altered on its journey to the photosphere, and hence we may not find the scalings we expect. Observations often only consider the line of sight magnetic field, the vertical magnetic field in our case, and so caution must be taken when making deductions about the interior magnetic field.

\begin{figure}[ht]
\centering
\begin{subfigure}{0.48\textwidth}
\centering
\includegraphics[scale=0.4]{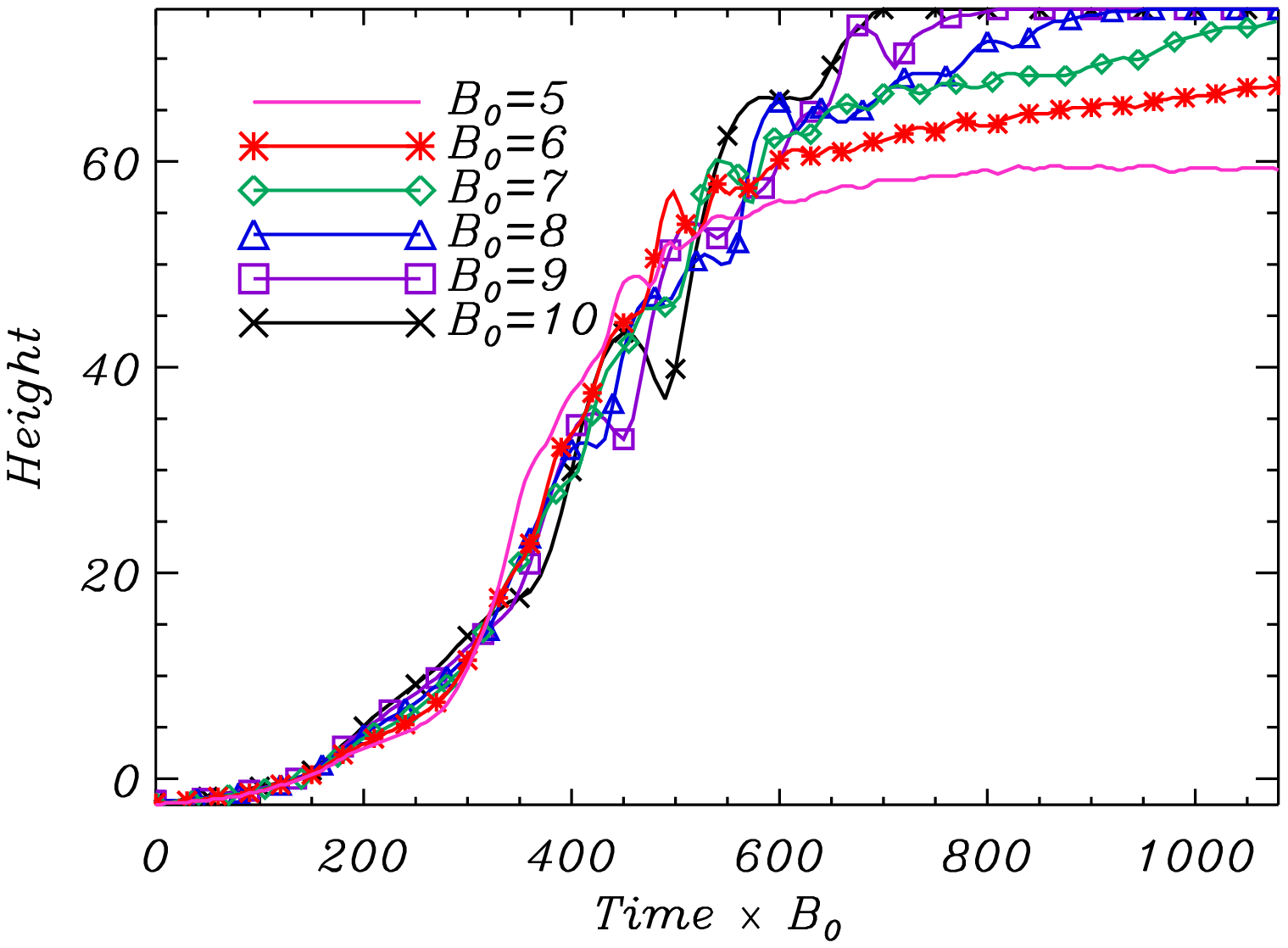}
\subcaption{}
\label{subfig:zfrontb0}
\end{subfigure}\\
\begin{subfigure}{0.48\textwidth}
\centering
\includegraphics[scale=0.4]{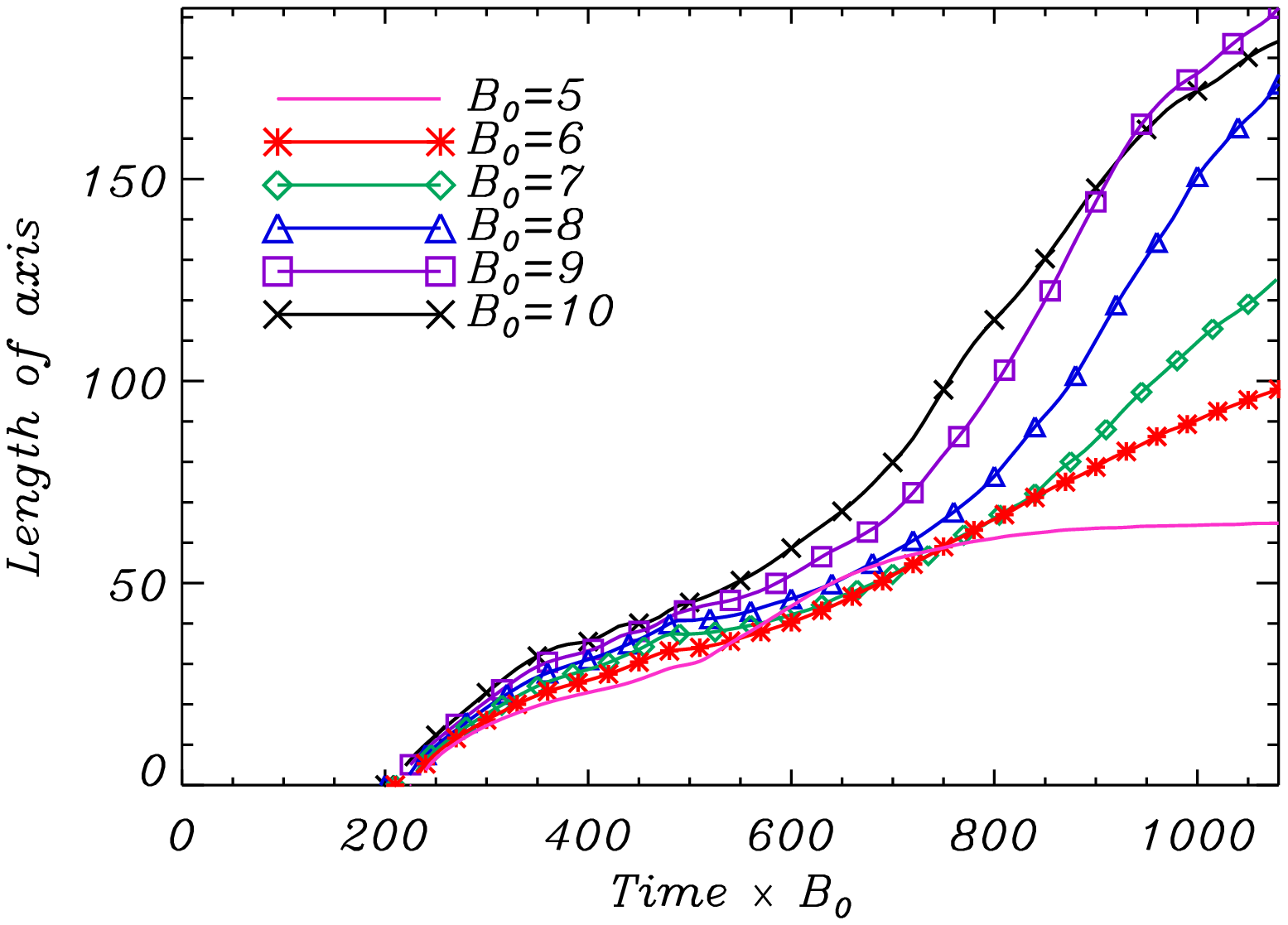}
\subcaption{}
\label{subfig:axislengthb0}
\end{subfigure}
\label{fig:generalrot2}
\caption{The evolution of (a) the height of the leading edge of the flux system and (b) the length of the axis fieldline in the atmosphere above $z=0$ for the varying $B_0$ over rescaled time.}
\end{figure}

In order to investigate the magnetic field's journey to the atmosphere, we also consider the leading edge of the flux system over rescaled time for the Group 1 cases in Figure~\ref{subfig:zfrontb0}. We find the evolution to be self-similar until approximately $\bar{t}=500$. The $B_0=5$ and $B_0=6$ cases appear to plateau at a fixed height, not reaching the top of the box. The height that the flux tubes reach is determined by pressure balance on the boundary, i.e. where the total pressure within the tube equals the background gas pressure. Emergence slows for the weaker $B_0$ experiments and there is not enough magnetic pressure to push the boundary upward. Hence, the maximum height reached is lower for weaker field experiments. It is hard to determine where stronger fields reach their pressure balance boundary as they reach the top of the box during the experiment. Consequently, the fieldlines in the stronger $B_0$ experiments extend further into the atmosphere and so their axis is longer as shown in Figure~\ref{subfig:axislengthb0}. This plot shows the length of the axis fieldline as measured from the point the fieldline enters the $z=0$ plane to the point where it leaves through $z=0$ after passing through the atmosphere. The axis of the $B_0=5$ tube appears to plateau at a fixed length, much shorter than that of the stronger $B_0$ tubes given that they extend higher into the corona. The length of the axis fieldline has an approximately linear dependence on the initial field strength, $B_0$. We return to this concept later as it has important consequences when calculating the twist.

\subsection{Torque}
\begin{figure}[ht]
\centering
\includegraphics[scale=0.4]{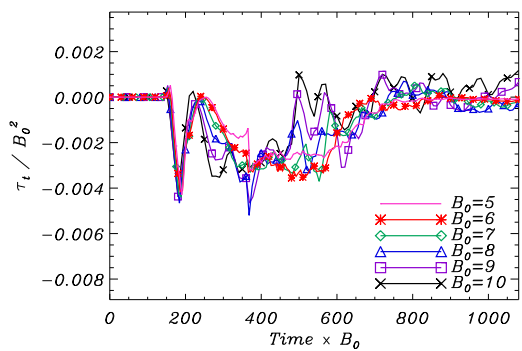}
\caption{Torque integrals due to magnetic tension for various $B_0$ cases (as defined in the key) with specifically the rescaled torque integrals, $\bar{\tau_T}=\tau_T/B_0^2$, measured over rescaled time, $\bar{t}=B_0t$. }
\label{fig:torqueb0}
\end{figure}

In Paper I, a full analysis was performed of the unbalanced torque caused by magnetic forces. This was concluded to be the driver of the rotational motion at the photosphere. The torque, $\tau_F$, is the tendency of a force to rotate an object about an axis and is given by $\mathbf{r} \times \mathbf{F}$ where $\mathbf{r}$ is the displacement vector from the centre and $\mathbf{F}$ is the force of interest. If we consider a closed circular curve surrounding the maximum of $B_z$ on the photospheric plane, and calculate a surface integral of the torque within this integral, we find the magnetic tension force to be the only contributor. Explicitly, we find the torque due to magnetic forces, $\tau_F$, to be equal to the torque due to magnetic tension, $\tau_t$, as follows
 \begin{equation*}
 \tau_F  = \tau_t = \iint{\mathbf{r} \times \left((\mathbf{B} \cdot \nabla) \mathbf{B}\right)   \cdot \mathrm{d} \mathrm{\bf S}},\nonumber
 \end{equation*}
 where $S$ is the surface contained within a circular contour of radius $a=2.5$ surrounding the maximum of $B_z$. Hence, we speculate that the unbalanced torque produced by the magnetic tension force drives the rotation. This has been plotted in Figure~\ref{fig:torqueb0} for all of the Group $1$ cases where we have rescaled both quantities as follows. We have again rescaled time as $\bar{t}=B_0t$ and we also redefine the torque integral by scaling it with respect to $B_0^2$ as $\bar{\tau_T} = \tau_T/B_0^2$, given the magnetic tension force is proportional to $B_0^2$.

As stated earlier, all experiments have been executed for different final times, in order to ensure they have the same final $\bar{t}$. This should ensure that we capture the same period of evolution for all of the varying $B_0$ experiments and that we are not missing some phases of the evolution for lower $B_0$ values. Generally, it appears that all six simulations demonstrate a self-similar torque imbalance which drives the rotation. The largest magnitude of torque is found between $\bar{t}=200$ and $\bar{t}=700$, after which the torque damps suggesting a slowing of the rotation after this point.

\subsection{Rotation angle}
\label{sec:rotangleb0}
\begin{figure}[ht]
\centering
\begin{subfigure}{0.48\textwidth} 
\centering
\includegraphics[scale=0.4]{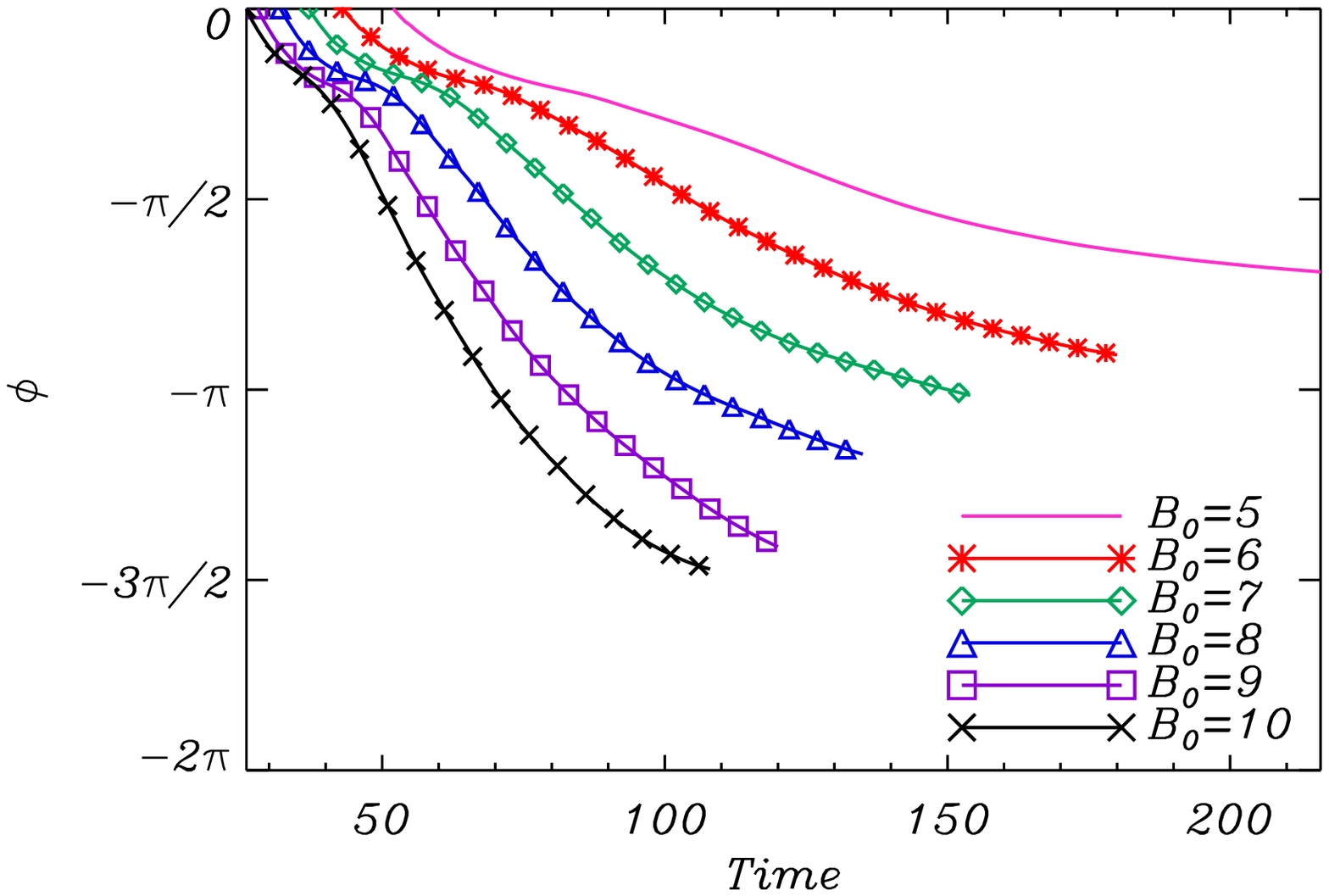}
\subcaption{}
\label{subfig:rotangleb0}
\end{subfigure}\\
\begin{subfigure}{0.48\textwidth}
\centering
\includegraphics[scale=0.4]{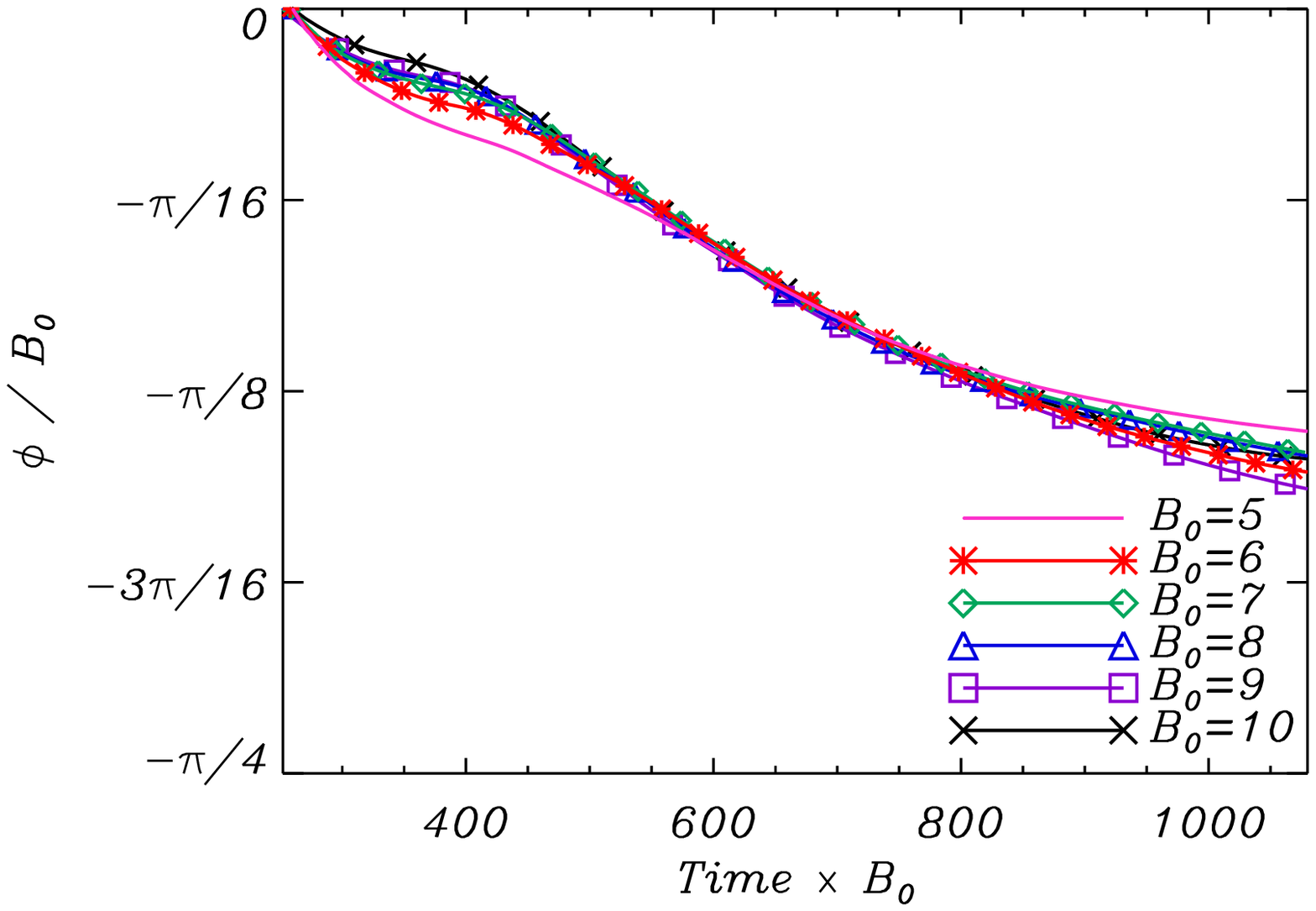}
\subcaption{}
\label{subfig:rotanglescaledb0}
\end{subfigure}\\
\begin{subfigure}{0.48\textwidth}
\centering
\includegraphics[scale=0.4]{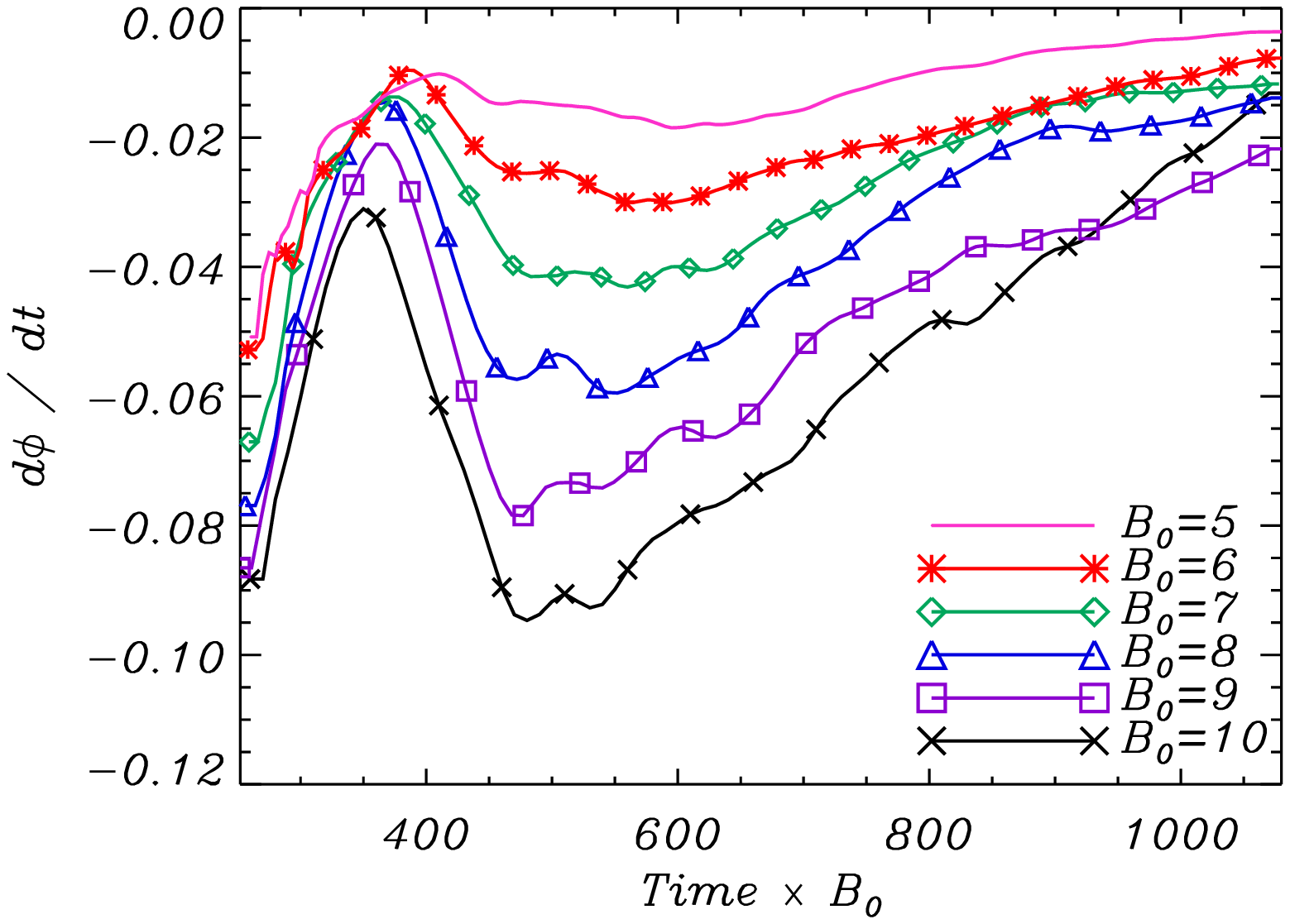}
\subcaption{}
\label{subfig:rotanglederivative}
\end{subfigure}
\caption{Rotation angles for various $B_0$ cases with (a) the unscaled rotation angles measured over time, (b) the rescaled rotation angles, $\bar{\phi}=\phi/B_0$, measured over rescaled time, $\bar{t}=B_0t$ and, (c) the unscaled rotation rate, $\mathrm{d}\phi/\mathrm{d}t$ measured over rescaled time. }
\label{fig:rotanglesb0}
\end{figure}

As discussed in Section~\ref{sec:general}, both sunspots experienced significant rotations in the general experiment. This is true for all $B_0$ cases to varying degrees. In order to calculate the rotation angle, we trace the photospheric location of a series of fieldlines using a fourth-order Runge-Kutta method from the base of the computational domain. In particular, we trace the axis fieldline from the centre of the negative flux source at $(0,-15,-25)$ and, as we expect, it follows the centre of the sunspot. In order to calculate the rotation angle, we also trace a selection of fieldlines from the base within a radius of one around the axis. Given the $x$ and $y$ coordinates of the intersections of selected fieldlines through the photosphere, we can calculate the angle of rotation using
\begin{equation*}
\phi=\tan^{-1}\left(\frac{y-y_{\text{axis}}}{x-x_{\text{axis}}}\right),
\end{equation*}
where $(x_{\text{axis}},y_{\text{axis}})$ is the location of the axis and $(x,y)$ is the location of another fieldline we have traced. To calculate the angle, we trace $100$ fieldlines from a circular footpoint of radius one on the base. We can then calculate the mean rotation angle by averaging the rotation angle over the traced fieldlines within the footpoint of interest. However, as all traced fieldlines intersect at different locations on the photosphere, and hence have different initial rotation angles, we must first subtract off the initial rotation angle, and therefore the average begins at $\phi=0$ for all cases. The resulting mean rotation angles are shown in Figure~\ref{subfig:rotangleb0} for the six cases we are investigating. 

After the emergence of the fields at the photosphere, there is a short period with little change in rotation angle while the sunspots drift apart. From Equation~\eqref{eq:densitydef}, the buoyancy force is proportional to $B_0^2$ and so flux tubes with larger $B_0$ values appear at the photosphere first. Consequently, the time taken for the flux tubes to reach the photosphere is inversely proportional to $B_0$. To incorporate this, we again redefine time as $\bar{t}=B_0t$ and rescale the horizontal axis as shown in Figure~\ref{subfig:rotanglescaledb0}. We also notice a direct relationship between $\phi$ and $B_0$ so we redefine $\bar{\phi}=\phi/B_0$ and find that the scaled rotation angles are approximately similar to each other. The scaled time evolution of the scaled mean rotation angles is shown in Figure~\ref{subfig:rotanglescaledb0}.   

\begin{table}[ht]
\centering
\caption{Rotation angles for Group $1$ experiments.}
\label{table:angleb0}
\begin{tabular}{ccccc}
\hline\hline
 \hspace{0.25cm} $B_0$ \hspace{0.25cm} & \hspace{0.25cm} $t$ \hspace{0.25cm} & \hspace{0.25cm} $\bar{t}$ \hspace{0.25cm} & \hspace{0.25cm} $\phi$ \hspace{0.25cm} & \hspace{0.25cm} $\phi/B_0$ \hspace{0.25cm} \\ [0.5ex]
 \hline\hline
 $5$ & $216$ & $1080$ & $124\degree$ & $25\degree$\\[0.5ex]
 $6$ & $180$ & $1080$ & $163\degree$ & $27\degree$   \\[0.5ex]
 $7$ & $154$ & $1080$ & $183\degree$ & $26\degree$ \\[0.5ex]
 $8$ & $135$ & $1080$ & $211\degree$ & $26\degree$\\[0.5ex]
 $9$ & $120$ & $1080$ & $254\degree$ & $28\degree$ \\[0.5ex]
 $10$ & $108$ & $1080$ & $265\degree$ & $26\degree$ \\[0.5ex]
 \hline
\end{tabular}
\end{table}

In Table~\ref{table:angleb0}, we have selected the magnitude of the final angles of rotation for the various $B_0$ cases. The second column of the table contains the unscaled time, $t$, the third column the rescaled time, $\bar{t}$, the fourth the unscaled rotation angle, $\phi$, and the fifth the rescaled rotation angle, $\phi/B_0$ to take into account the rotation angle's dependence on the magnetic field strength. We have chosen to consider the rotation angles at the final rescaled time of $\bar{t}=1080$ as we expect the flux tubes to be in a similar stage of their evolution here. This is presented in the fifth column of the table and shows that the scaled rotation angles are approximately constant. A similar analysis has been conducted for the other sunspot with comparable results.

By rescaling $\phi$ with respect to $B_0$, the only varied quantity in this model, we are able to remove any dependency on $B_0$ and reveal a self-similar behaviour as the fieldlines threading the sunspot rotate around the centre fieldline. On first inspection, this result may seem surprising as initially all fieldlines have the same helical structure since the degree of twist, $\alpha$, is constant in this group. Hence, we may have expected all experiments to have the same final rotation angle. This suggests that varying the field strength not only affects the timing at which key processes occur but also the amount by which the fieldlines rotate. In particular, we believe that the length of the fieldlines extending into the atmosphere, as ordered by $B_0$, is influencing the rotation angle at the photosphere.

Furthermore, the rotation rate, $\mathrm{d}\phi/\mathrm{d}t$, as displayed in Figure~\ref{subfig:rotanglederivative}, drops off towards the end of all experiments, suggesting that the rotation may not significantly persist if the experiments were continued. By demonstrating the rotation angle's dependence on the field strength, this corroborates the theory introduced by~\cite{Min2009} that the rotation is a consequence of the torque on the photospheric boundary rather than by apparent effects. If the rotation was due to apparent effects, altering the field strength of the flux tube would not vary the amount of rotation. The relationship between the rotation angle and $B_0$, and the ramifications of this are investigated in further detail in later sections.

\subsection{Twist}
In order to estimate the twist of the magnetic field, we investigate a number of twist related quantities. To begin, we calculate the twist of individual fieldlines. Precisely, we consider the fieldline twist of $100$ fieldlines stemming from a footpoint of radius one surrounding the axis of the sunspot. This is the same approach we used when calculating the rotation angle. To determine the twist of a fieldline we calculate
\begin{equation}
\Psi = \int{\mathrm{d}\psi} = \int_0^L{\frac{B_{\psi}}{RB_{n}}}~\mathrm{d}l\nonumber
\end{equation}
along the length of a fieldline, $L$, in a local cylindrical coordinate system $(R,\psi,n)$. Consider a distance of one unit along the axis, henceforth referred to as an axial unit length. Then, $B_{\psi}/(RB_{n})$ is the angle through which the fieldlines rotate over one axial unit length. Hence, by summing this quantity over the length of the axis fieldline, we calculate the angle in radians through which fieldlines rotate over the axial length. We can then deduce the number of turns, $N$, the fieldlines pass through over the axial length by noting $\Psi=2\pi N$.

To transform the coordinate system from the original Cartesian system to this local cylindrical system, we trace the axial fieldline using a fourth-order Runge-Kutta method and define a plane at each step along the axis fieldline where the normal to the plane, $\mathbf{\hat{n}}$, is directed along the axis. Next, we create a local cylindrical system within each of these planes by defining $R$ and $\psi$ as the respective radial and azimuthal directions within the plane. Next, the outer fieldlines are traced and the angle through which they rotate is calculated by summing over the angle found when they intersect each of the planes.

\begin{figure}[!ht]
\centering
\includegraphics[scale=0.4]{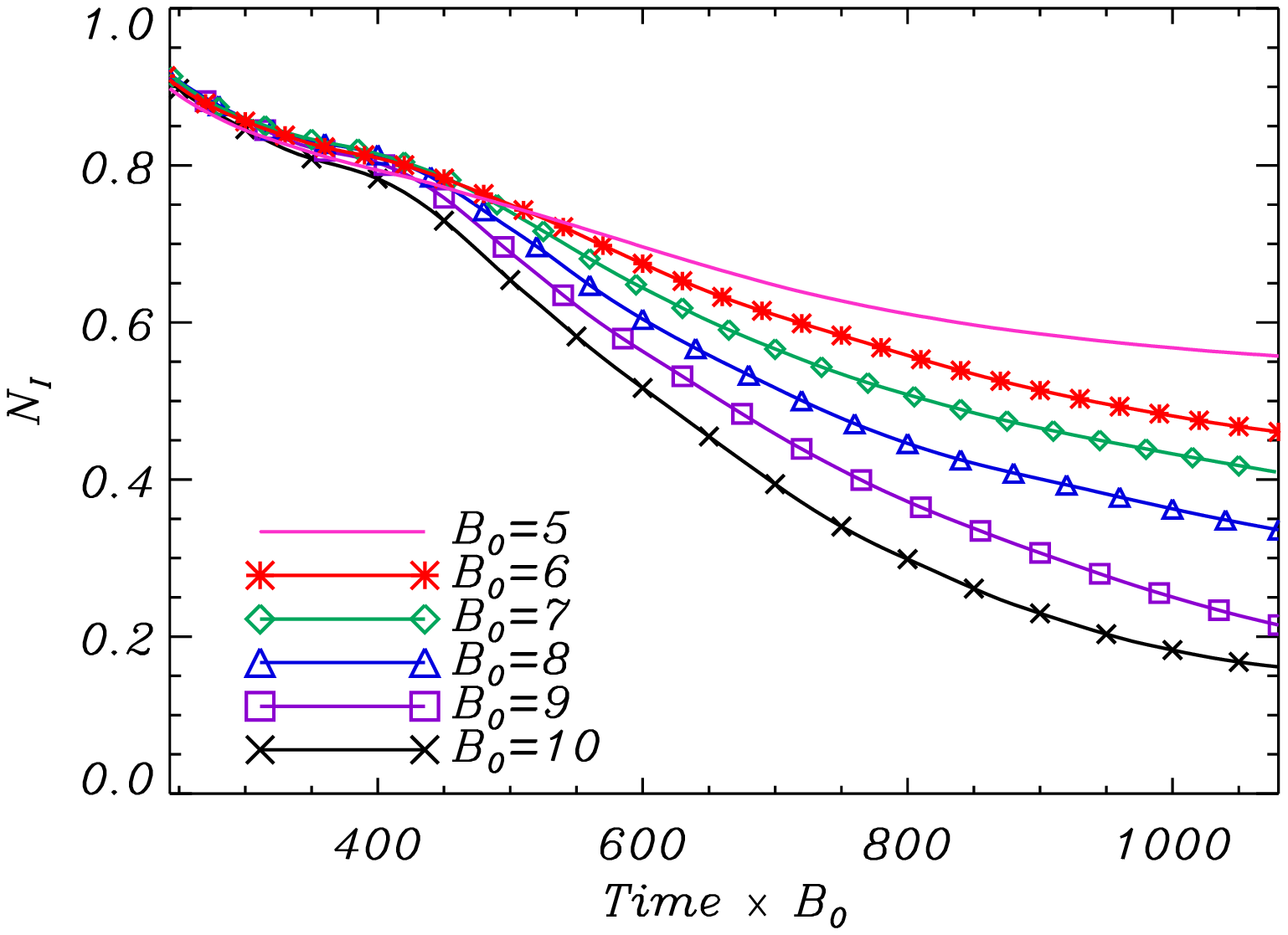}
\caption{Average number of turns, $N_{\text{I}}$, fieldlines undergo within the interior portion of one leg of the tube, measured over rescaled time, $\bar{t}=B_0t$ for Group 1 cases.}
\label{fig:twistb0}
\end{figure}

Using the method described above, the fieldline twist, and hence the number of turns the fieldlines pass through, is calculated for $100$ fieldlines originating from the left footpoint within a radius of one. To gain a mean value, we average over the number of turns the fieldlines pass through for all fieldlines traced. This is plotted against scaled time in Figure~\ref{fig:twistb0}. More precisely, we have plotted the average number of turns  fieldlines pass through within an interior section of a leg of the flux tube, i.e. up until $z=0$, by averaging,
\begin{equation}
N_{\text{I}} = \frac{\Psi_{\text{I}}}{2\pi} = \frac{1}{2\pi}\int_{y<0,~z<0}{\frac{B_{\psi}}{RB_{n}}}~\mathrm{d}l,\label{eq:turnsinterior}
\end{equation}
over $100$ fieldlines.

Notice, we are again measuring this quantity over rescaled time to take into account the time lag associated with weaker initial field strengths. Initially all fields have the same number of turns around the axis within the interior as they contain the same initial twist. The initial evolution is similar for all cases but the twist drops off more sharply for higher $B_0$ cases. This result may seem surprising due to the initial twist profile. However, this discrepancy is likely to be related to the expansion of the field in the atmosphere. The stronger $B_0$ cases expand higher into the atmosphere distributing the atmospheric twist along a longer length (see Figure~\ref{subfig:axislengthb0}) resulting in a smaller twist per unit length in the atmospheric portion of the field. This produces a larger gradient in the twist per unit length, driving the rotation and untwisting the interior field. Notice, there are large amounts of residual twist in the submerged legs of the tube for the weaker field cases. This is related to the distribution of the twist per unit length across the domain.

Unfortunately, we cannot accurately calculate the fieldline twist within the atmospheric section of the tube as the axis kinks from the transverse $x$ direction and the assumptions we require to change into our local planar coordinate system are no longer valid. Hence, this has been excluded from our discussion of the fieldline twist.

\begin{figure}[!ht]
\centering
\begin{subfigure}{0.48\textwidth}
\includegraphics[scale=0.4]{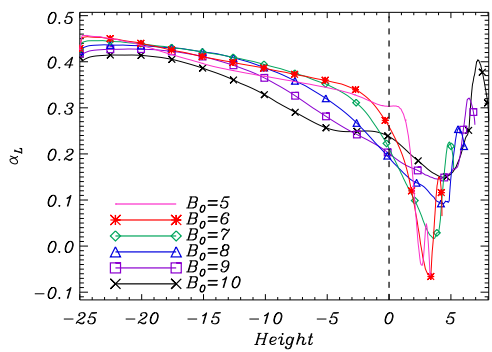}
\subcaption{$\bar{t}=450$}
\label{subfig:alphaalongaxisb0a}
\end{subfigure}\\
\begin{subfigure}{0.48\textwidth}
\includegraphics[scale=0.4]{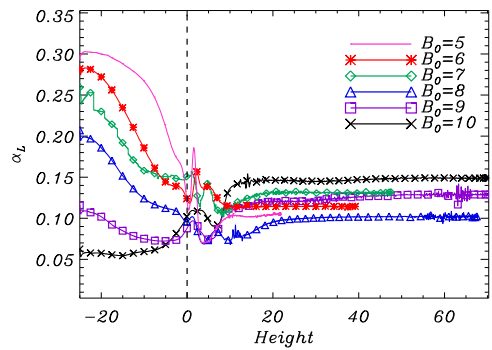}
\subcaption{$\bar{t}=1080$}
\label{subfig:alphaalongaxisb0b}
\end{subfigure}
\caption{The quantity $\alpha_L$ traced along the axis fieldline against height, $z$, of the axis for two scaled times, (a) $\bar{t}=450$ and (b) $\bar{t}=1080$. The dashed black line denotes the height at the solar surface, $z=0$.}
\label{fig:alphaalongaxisb0}
\end{figure}

In an attempt to explain the large amounts of twist left in the interior for lower $B_0$ experiments, we present an analysis of a proxy for the local twist referred to as the force-free parameter, defined as 
\begin{equation*}
\alpha_L = (\mathbf{\nabla} \times \mathbf{B}) \cdot \mathbf{B} / B^2.
\end{equation*}
This quantity has been shown to be closely linked to the twist per unit length and has been calculated as such in many previous studies, for example~\cite{Fan2009} and~\cite{Sturrock2015}. The parameter is also used as a measure of twist in observational studies (see~\citealp{Liu2014} or~\citealp{Hahn2005}). It should be noted that this is not the $\alpha$ we vary in Group $2$ and we have denoted this as $\alpha_L$ to differentiate between the two. If we assume $\alpha_L$ to be constant and assume the field is force-free, $\alpha_L$ can be shown to be equal to twice the twist per unit length. In order to visualise the distribution of $\alpha_L$ along different sections of the tube, we have traced $\alpha_L$ along the axis passing through the centre of the sunspot from the left footpoint to the apex of the fieldline. To try and compare the different $B_0$ cases, we have considered a snapshot of $\alpha_L$ along the axial fieldline at two scaled times, $\bar{t}=450$ and $\bar{t}=1080$ as shown in Figure~\ref{fig:alphaalongaxisb0}. Notice, we have plotted $\alpha_L$ against the height of the tube. Hence, the stronger $B_0$ tubes have reached further into the atmosphere by the final snapshot at $\bar{t}=1080$. In addition, it should be noted that we have plotted the symbols much less frequently due to the large number of steps we have taken along the axis fieldline.

Although not presented in this paper, at $t=0$, $\alpha_L$ is constant along the axis fieldline at a value of $0.6$ (twice the initial twist per unit length). However, this value drops as the field begins to expand into the atmosphere as shown in Figure~\ref{subfig:alphaalongaxisb0a} at $\bar{t}=450$ soon after the field has emerged. We see that the value of $\alpha_L$ drops off with increasing height for all experiments, and hence a gradient develops in $\alpha_L$.~\cite{Longcope2000} predicted that this gradient in $\alpha_L$ produces a torque (as we found earlier) that drives the torsional motion of the flux tube intersecting the photosphere. The authors hypothesised that the torsional motion will continue until this gradient in $\alpha_L$ is removed. 

At the final scaled time, $\bar{t}=1080$, we find that the axis of the flux tubes have reached higher into the atmosphere. The magnitude of $\alpha_L$ appears approximately constant in the coronal portion of the field, indicating that this section of the field is almost force-free. The low-magnitude in $\alpha_L$ likely arises because of the stretching and expansion of the field (see~\citealp{Fan2009}). Interestingly, the magnitude of $\alpha_L$ in the interior is ordered by $B_0$ such that stronger magnetic flux tubes process a lower magnitude of $\alpha_L$ and weaker magnetic flux tubes retain a higher magnitude of $\alpha_L$ in this region. As mentioned earlier, it is often conjectured that the cause of the rotation at the photosphere is related to $\alpha_L$ trying to equilibrate between the twisted interior and stretched coronal field. However, this is not yet the case in the weaker field experiments as a higher magnitude $\alpha_L$ persists in the interior at the final time. Furthermore, the strongest experiments ($B_0=9$ and $B_0=10$) appear to have distributed their twist to the stage where $\alpha_L$ is larger in the atmosphere. This suggests that the experiments have passed through the equilibrium state as the tubes have ``over-rotated''. The experiments would need to be performed for a longer time to determine whether the twist is equilibrating on a longer timescale or whether the lower strength field cases are unable to unwind their interior twist to match their coronal twist.

It is important to note that although Figure~\ref{subfig:alphaalongaxisb0b} is plotted at the same scaled time for all experiments, all experiments appear to be at slightly different stages in their evolution. The reasons behind this are two-fold. Firstly, the weaker fields are equilibrating on a shorter distance and secondly, the Alfv\'{e}n speed is also proportional to $1/\sqrt{\rho}$. Given the density deficit's dependence on $B_0$, the density deficit of the tube is non-linearly related to $B_0$ with weaker tubes possessing a larger density. This in turn means weaker tubes have a slightly smaller $v_A$, and as such may be evolving on a slower timescale than predicted by $\bar{t}=B_0t$. 

Additionally, it is worth noting that as the weaker fields do not extend as far into the atmosphere, the axis fieldline is shorter. If the twist per unit length, $\alpha_L$, equilibrates on a shorter fieldline, the average value of $\alpha_L$ will be significantly larger in both the interior and atmospheric regions. Explicitly, if we assume that $\alpha_L$ tends to a constant value and that the total twist is conserved, then $\alpha_il_i=\alpha_fl_f$ where $\alpha_i$ is the initial twist per unit length, $\alpha_f$ the final, $l_i$ the initial axial length, and $l_f$ the final. In this case the predicted final twist per unit length is $\alpha_f=\alpha_il_i/l_f$. Since all experiments have the same $\alpha_i$ and $l_i$ values, an increase in $l_f$ decreases the final twist per unit length, $\alpha_f$. In the weaker cases, the final rotation angle will be smaller as the tube does not need to unwind as much interior twist to achieve its larger final twist per unit length. This effect is not yet apparent but again may be seen if the experiments had been performed for a longer time.

\subsection{Vorticity}
\begin{figure}[!ht]
\centering
\begin{subfigure}{0.48\textwidth}
 \centering
\includegraphics[scale=0.4]{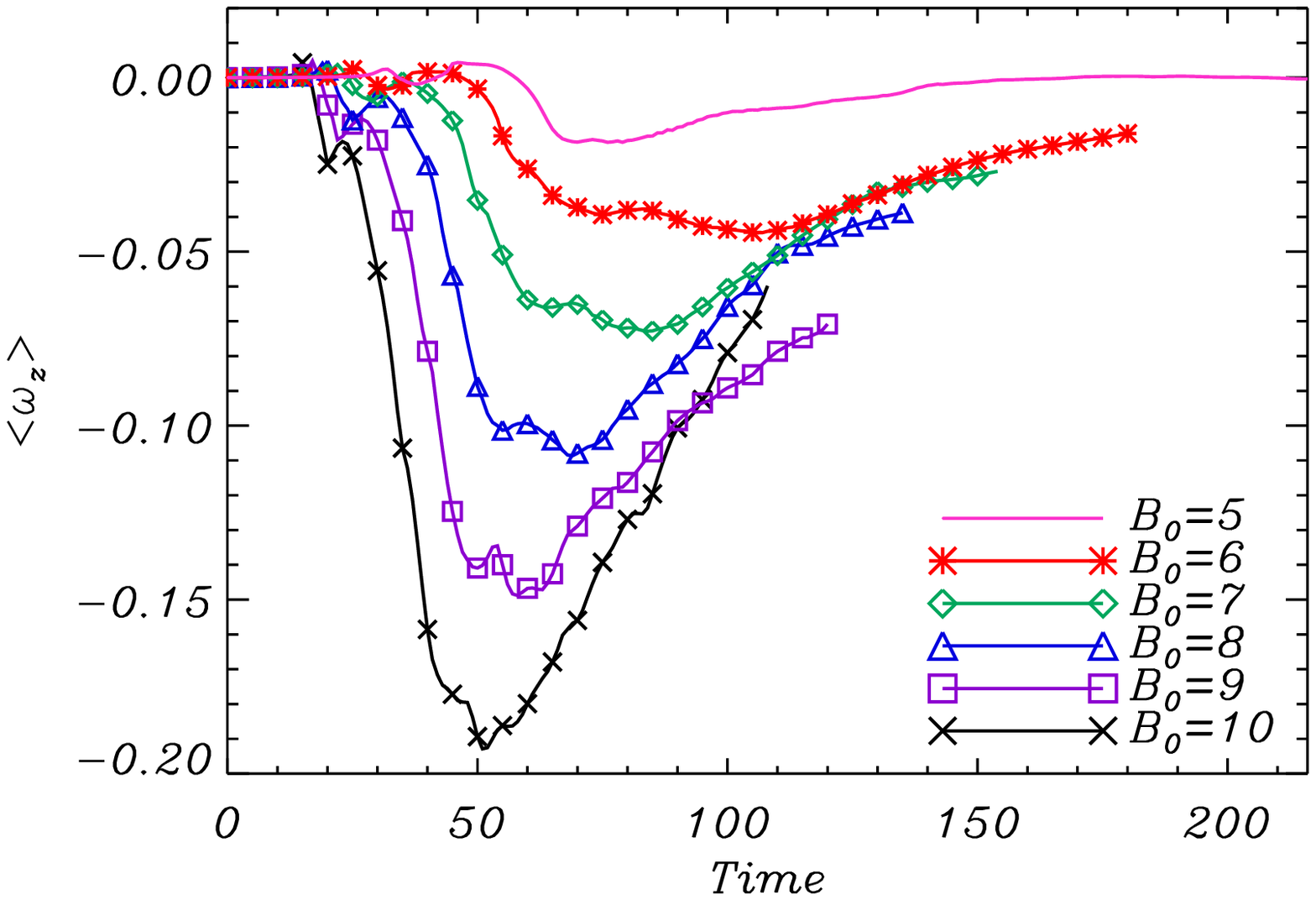}
\subcaption{}
\label{subfig:vorticityb0}
\end{subfigure}\\
\begin{subfigure}{0.48\textwidth} 
\centering
\includegraphics[scale=0.4]{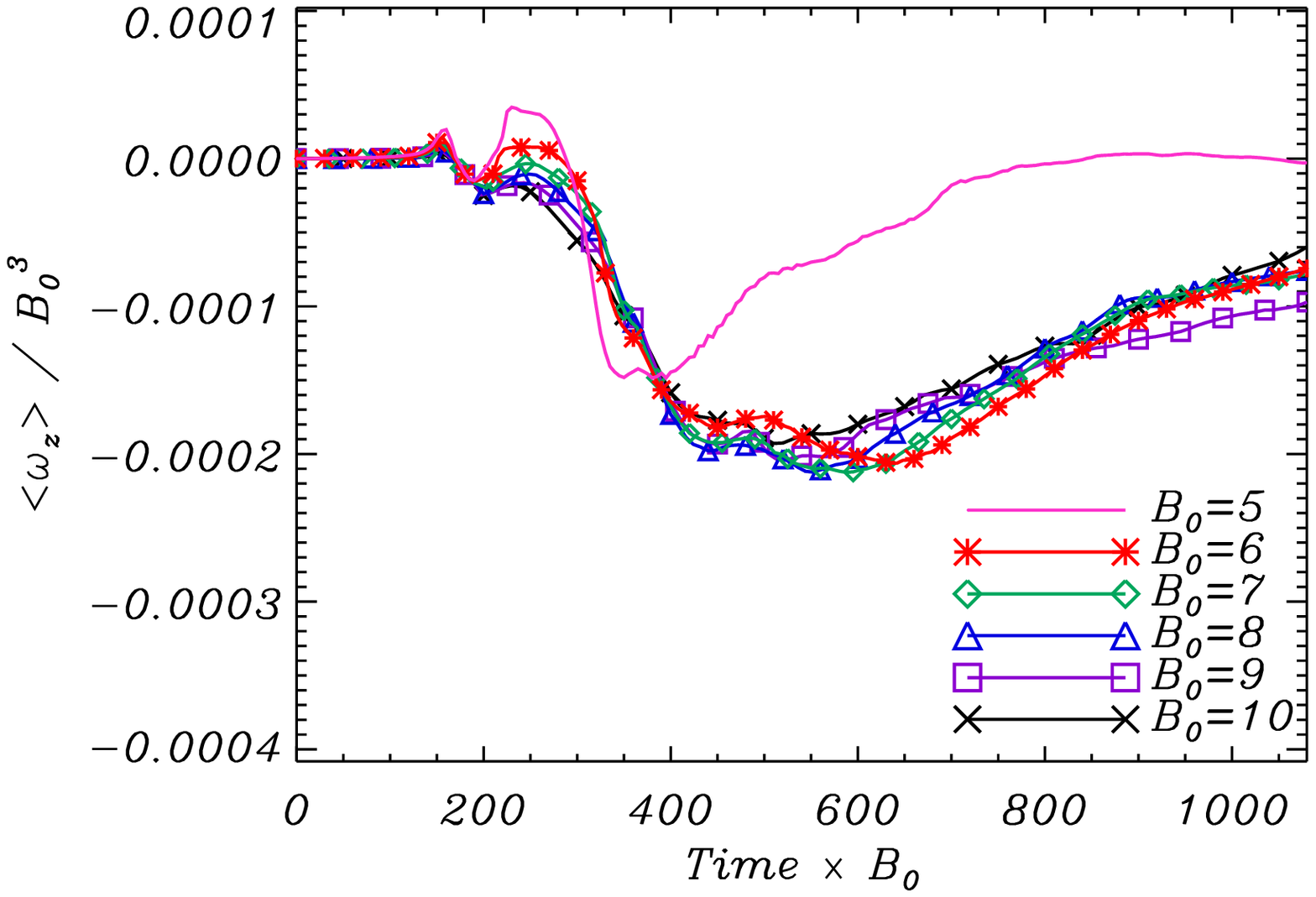}
\subcaption{}
\label{subfig:vorticityscaledb0}
\end{subfigure}
\caption{Average vorticity for Group 1 experiments with (a) the unscaled average vertical vorticity measure over time and (b) the rescaled average vorticity, $\bar{\langle{\omega_z}\rangle}=\langle\omega_z\rangle/B_0^3$, measured over rescaled time, $\bar{t}=B_0t$.}
\label{fig:vorticityb0}
\end{figure}

\noindent As discussed in Section~\ref{sec:rotangleb0}, all six simulations in Group $1$ exhibit rotation in their sunspots, quantifiable in terms of an angle. To examine this further, we have also calculated the mean vertical vorticity within each sunspot, as given by
\begin{equation*}
\langle \omega_z \rangle=\langle (\nabla \times \mathbf{v})_z \rangle=\left\langle \frac{\mathrm{d}v_y}{\mathrm{d}x} - \frac{\mathrm{d}v_x}{\mathrm{d}y}\right\rangle,
\end{equation*}
where we have averaged over the photospheric region where $B_z > 3/4\text{max}(B_z)$. This quantifies the rotation of the plasma within the upper polarity source. This has been plotted for all six Group $1$ experiments as shown in Figure~\ref{subfig:vorticityb0}. 

The average vertical vorticity is consistently negative for all $B_0$ cases indicating that the dominant motion within the sunspots is a clockwise rotation, consistent with the theory suggested in Paper I. Precisely, this rotation acts to untwist the interior magnetic field and inject twist into the atmospheric field. A very clear trend develops in that tubes with a stronger initial field strength emerge more quickly and significant vortical motions develop within their sunspot centres. To try and explore the relationship between $\langle \omega_z \rangle$ and $B_0$, we have rescaled $\langle \omega_z \rangle$ with respect to $B_0^3$ as well as redefining the time as $\bar{t}=B_0t$. The rescaled plot is shown in Figure~\ref{subfig:vorticityscaledb0} where again we find a self-similar evolution, apart from during the latter stages of the $B_0=5$ case. The difference in this case likely arises because we are not capturing the correct area within which the vorticity lies. In the weak $B_0=5$ case, the field spreads over the photosphere in order to build up enough field strength to initiate the magnetic buoyancy instability. This is not captured when considering the area where $B_z>3/4~\text{max}B_z$. The $B_0^3$ scaling is surprising and may be related to how we calculate $\langle{\omega_z}\rangle$. Future studies should repeat this to see if this is a viable trend. In this particular case, the scaling itself is not of great significance. More importantly, we should conclude that stronger $B_0$ tubes tend to have stronger vortical motions developing in their polarity sources.

\subsection{Current density}
\label{sec:currentb0}
Next, we consider another estimate for the twist of the magnetic field by calculating the electric current density, specifically the z-component,
\begin{equation*}
j_z= \frac{\partial B_y}{\partial x} - \frac{\partial B_x}{\partial y},
\end{equation*}
as this is linked to how twisted the magnetic field is in the photospheric $x-y$ plane. In a similar manner to Paper I, we analyse the vertical current density at two different planes, the photospheric plane, $z=0$, and a plane at the centre of the interior domain, $z=-12.5$.
\begin{figure}[!ht]
\centering
\begin{subfigure}{0.48\textwidth}
\centering
\includegraphics[scale=0.4]{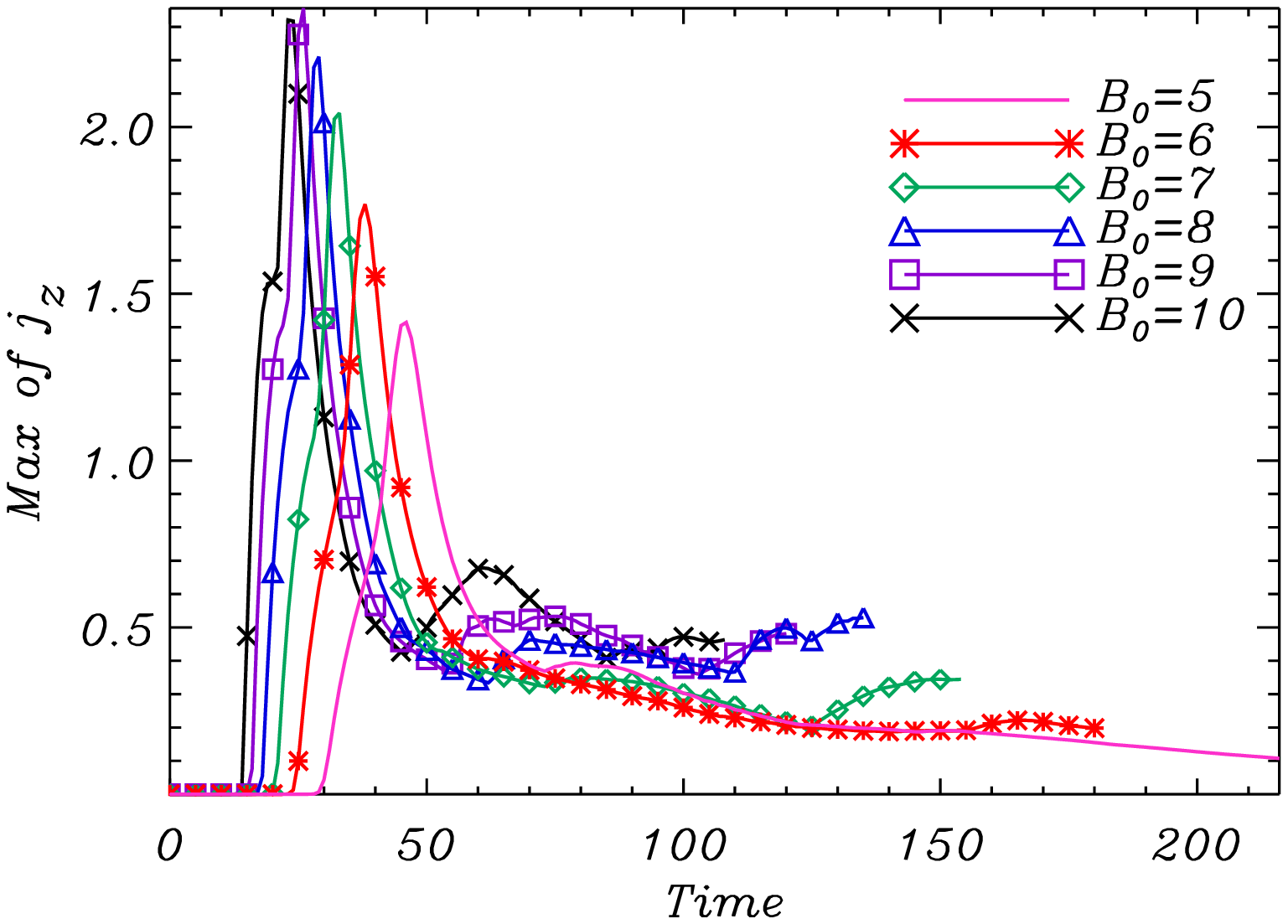}
\subcaption{}
\label{subfig:currentz0b0}
\end{subfigure}\\
\begin{subfigure}{0.48\textwidth}
\centering
\includegraphics[scale=0.4]{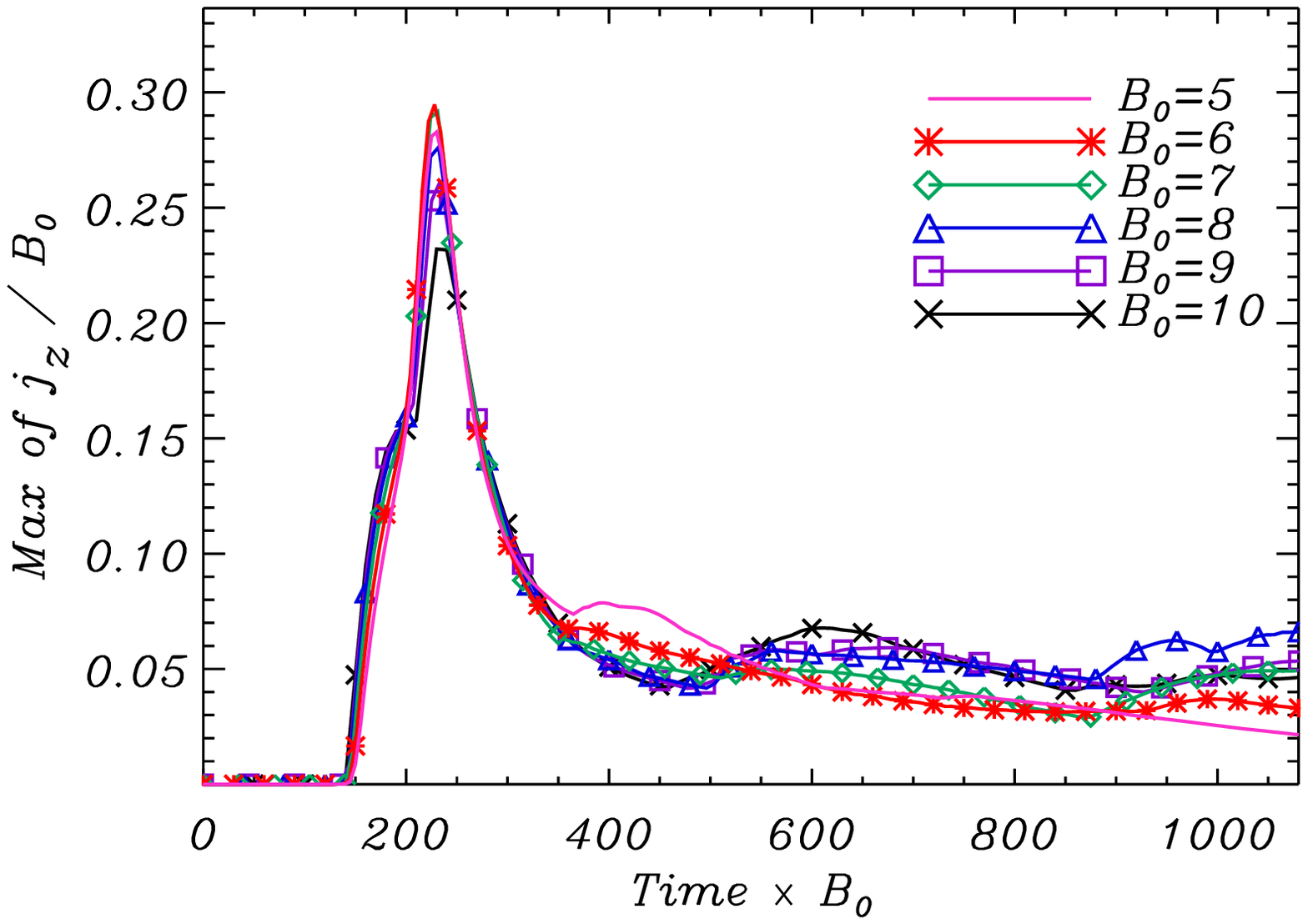}
\subcaption{}
\label{subfig:currentz0b0scaled}
\end{subfigure}\\
\begin{subfigure}{0.48\textwidth}
\centering
\includegraphics[scale=0.4]{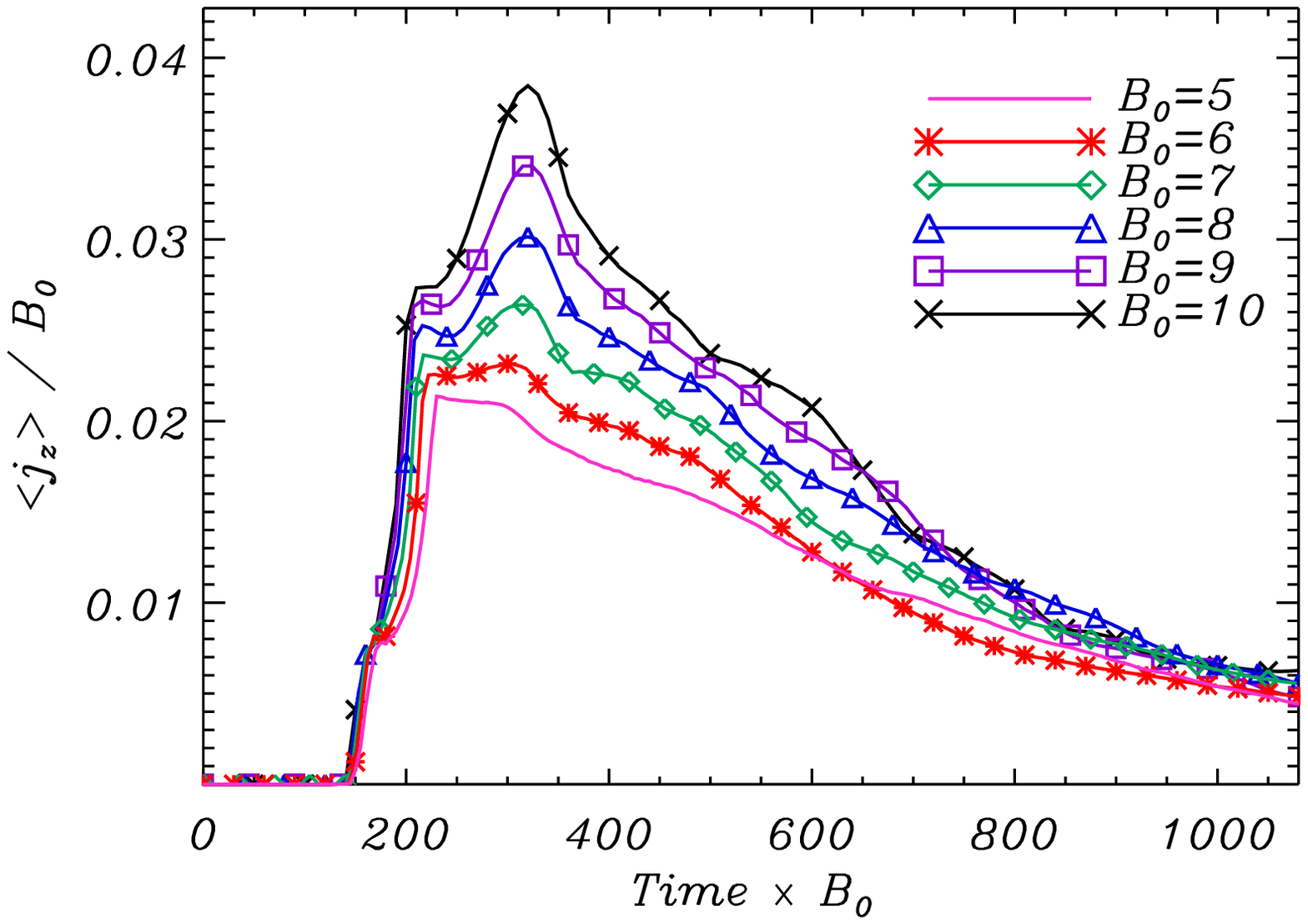}
\caption{}
\label{subfig:averagecurrentz0b0scaled}
\end{subfigure}
\caption{Maximum of $j_z$ over photospheric $z=0$ plane for various $B_0$ cases with (a) the unscaled maximum of $j_z$ measured over time 
and (b) the scaled maximum of $j_z$, $\text{Max of }j_z/B_0$ measured over rescaled time, $\bar{t}=B_0t$. Additionally, (c) shows the scaled average of $j_z$, $\langle{j_z\rangle}$, measured over rescaled time. }
\label{fig:currentz0b0}
\end{figure}
There are several proxies we can consider for measuring the current. In this case, we plot the temporal evolution of the maximum of $j_z$ for the $z=0$ plane in Figure~\ref{subfig:currentz0b0}. All cases show an initial peak in the maximum of $j_z$ at the photosphere due to the emergence of the field. The timing of this maximum is clearly dependent on the value of $B_0$ as the emergence timescale is proportional to $B_0$. The magnitude of the peak is also dependent on $B_0$ as we find the peak in the curve to be higher for larger $B_0$ values. This is as we expected given that $j_z$ is proportional to $B_0$. Later, all plots show a steady decline due to the expansion of the field into the higher atmosphere. To investigate the self-similarity in this plot, we have rescaled the maximum of $j_z$ with respect to $B_0$ and also rescaled time to become an Alfv\'{e}n time as discussed before. The result of the rescaling is shown in Figure~\ref{subfig:currentz0b0scaled} where we find a clear self-similarity in the curves as they lie on top of one another. From Figure~\ref{subfig:averagecurrentz0b0scaled}, however, we note that the average of $j_z$ does not directly scale with $B_0$. Stronger field experiments have larger averaged currents than predicted by the $B_0$ scaling, perhaps due to the larger rotation angle and vortical motions seen for stronger experiments.

\begin{figure}[!ht]
\centering
\begin{subfigure}{0.48\textwidth}
\centering
\includegraphics[scale=0.4]{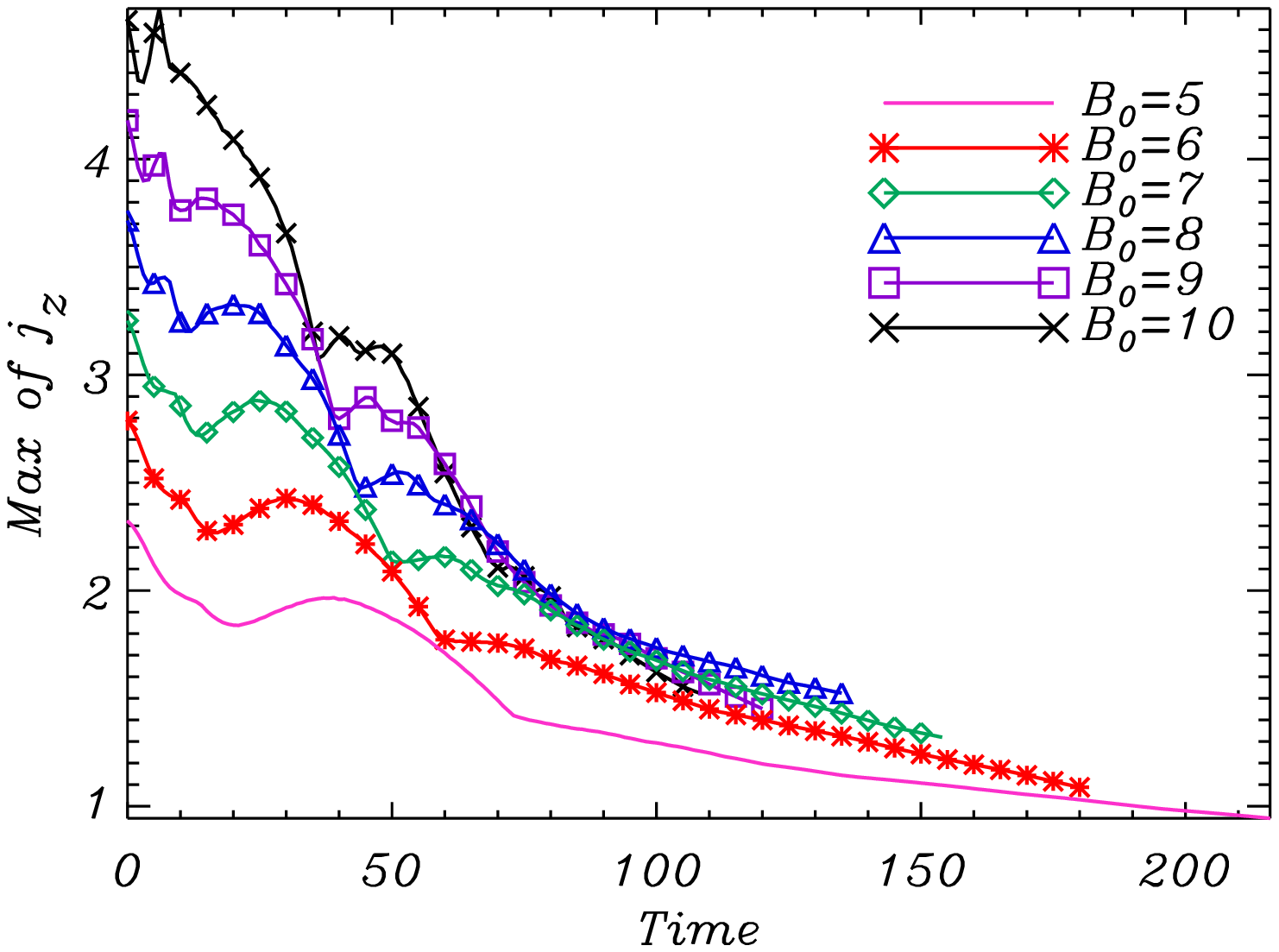}
\subcaption{}
\label{subfig:currentz-12b0}
\end{subfigure}\\
\begin{subfigure}{0.48\textwidth}
\centering
\includegraphics[scale=0.4]{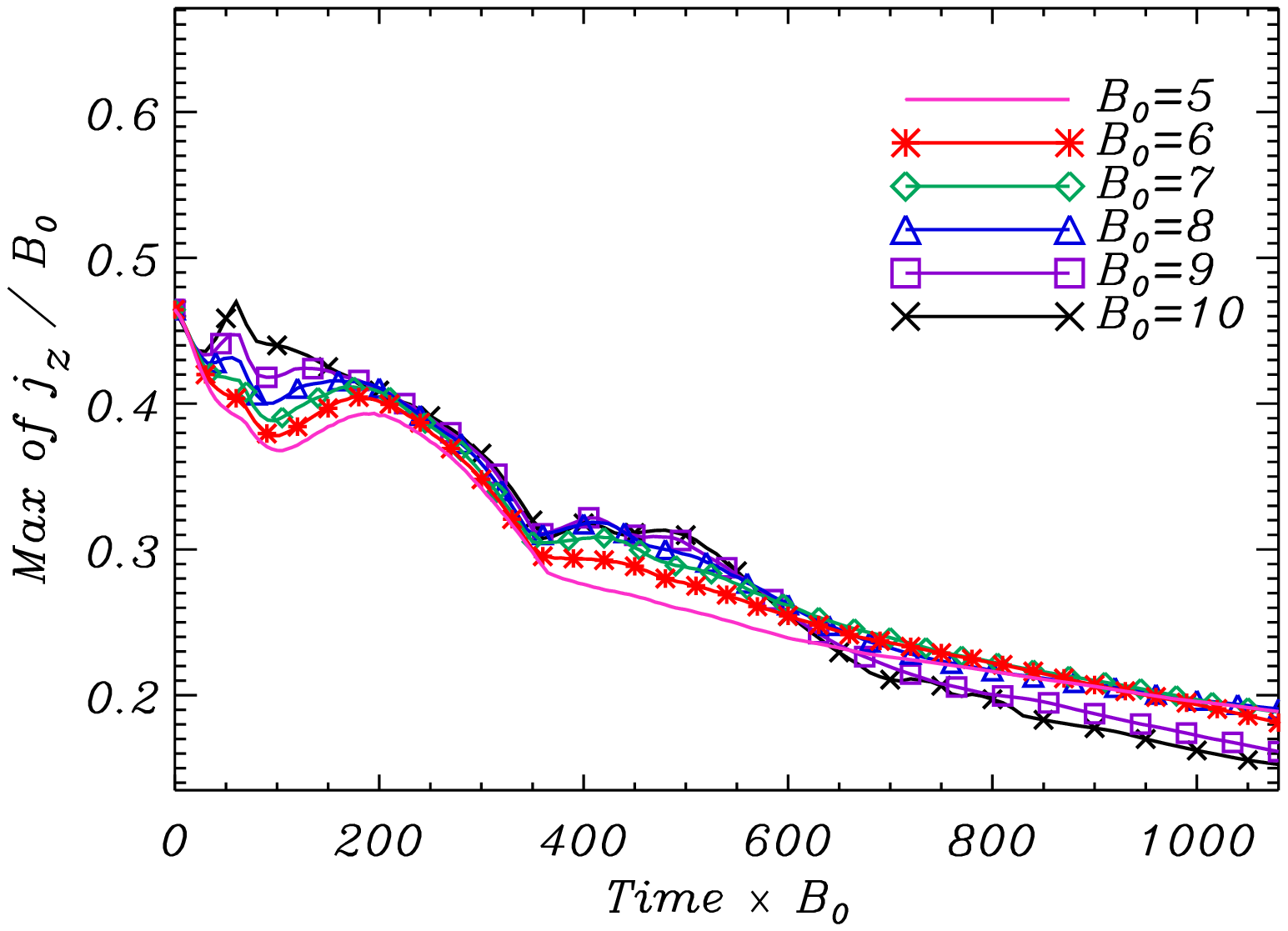}
\subcaption{}
\label{subfig:currentz-12b0scaled}
\end{subfigure}
\caption{Maximum of $j_z$ over the interior $z=-12.5$ plane for various $B_0$ cases with (a) the unscaled maximum of $j_z$ measured over time and (b) the scaled maximum of $j_z$, $\text{Max of }j_z/B_0$, measured over rescaled time, $\bar{t}=B_0t$.}
\label{fig:currentz-12b0}
\end{figure}

Similarly, we have plotted the spatial maximum of $j_z$ half way down the solar interior at $z=-12.5$, as displayed in Figure~\ref{subfig:currentz-12b0}. In all cases, after an initial decrease in the maximum of $j_z$, there is a slight increase due to the straightening of the legs of the tube. However, later there is a steady decrease in the maximum of $j_z$. Again, we find the timing of the maximum to be dependent on the initial field strength $B_0$. To take this into account, we have again rescaled the current and time to produce Figure~\ref{subfig:currentz-12b0scaled}. A clear self-similarity is seen here.

\subsection{Magnetic helicity}
To investigate the distribution of twist across the domain, we analyse the relative magnetic helicity within different sub-volumes of the domain. This quantitatively describes the degree of twist and shear of magnetic fieldlines. The share of magnetic helicity throughout the domain is affected by a number of factors, but mainly by vertical flows that move twisted magnetic fields into the corona and horizontal flows at the photosphere that shear and twist up magnetic fields~\citep{Berger1984}.

The magnetic helicity of the magnetic field $\mathbf{B}$, relative to some potential field $\mathbf{B}_p$, in a volume $V$ is defined as~\citep{Berger1984,Finn1985}
\begin{equation}
H_{r}= \int_{V}{(\mathbf{A}+\mathbf{A}_{p}) \cdot (\mathbf{B} - \mathbf{B}_p)\text{ }\mathrm{d}V},
\label{eq:helicity}
\end{equation}
where $\mathbf{A}$ is the vector potential of $\mathbf{B}$ and $\mathbf{A}_p$ is the vector potential of $\mathbf{B}_p$ such that $\mathbf{B}_p$ is the reference potential field with the same normal flux distribution as $\mathbf{B}$ on all bounding surfaces of $V$. This form of helicity is chosen as it is independent with respect to the choice of gauge $\mathbf{A}$. To calculate this quantity numerically within different sub-volumes, we use a numerical procedure first introduced in~\cite{Moraitis2014}. For further details of the method used to calculate the magnetic potential and vector potentials see~\cite{Valori2012}, and for details of how we implement the code in our particular experiments see Paper I.

\begin{figure}[!ht]
\centering
\begin{subfigure}{0.48\textwidth}
\centering
\includegraphics[scale=0.4]{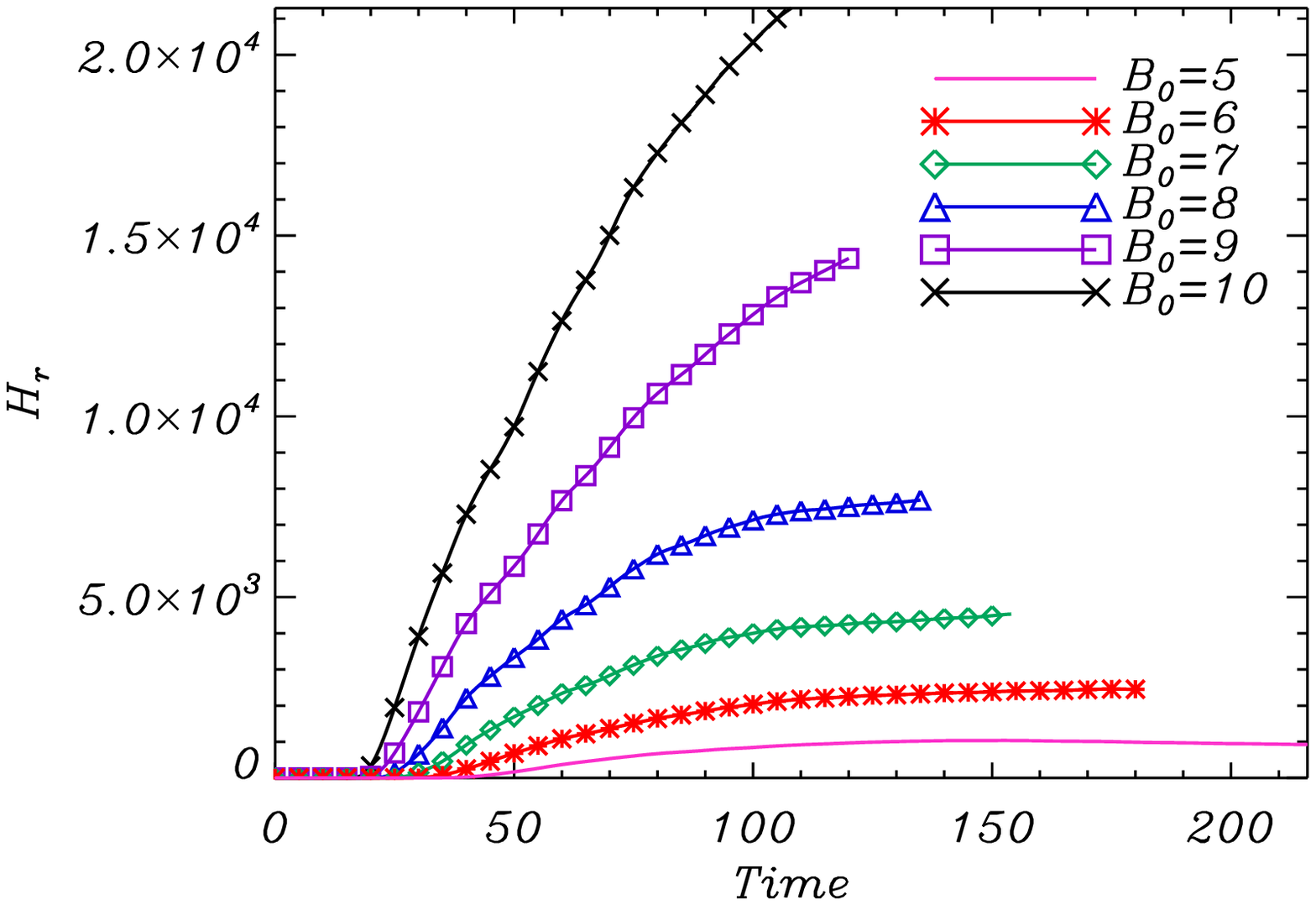}
\subcaption{}
\label{subfig:atmoshelicityb0}
\end{subfigure}\\
\begin{subfigure}{0.48\textwidth}
\centering
\includegraphics[scale=0.4]{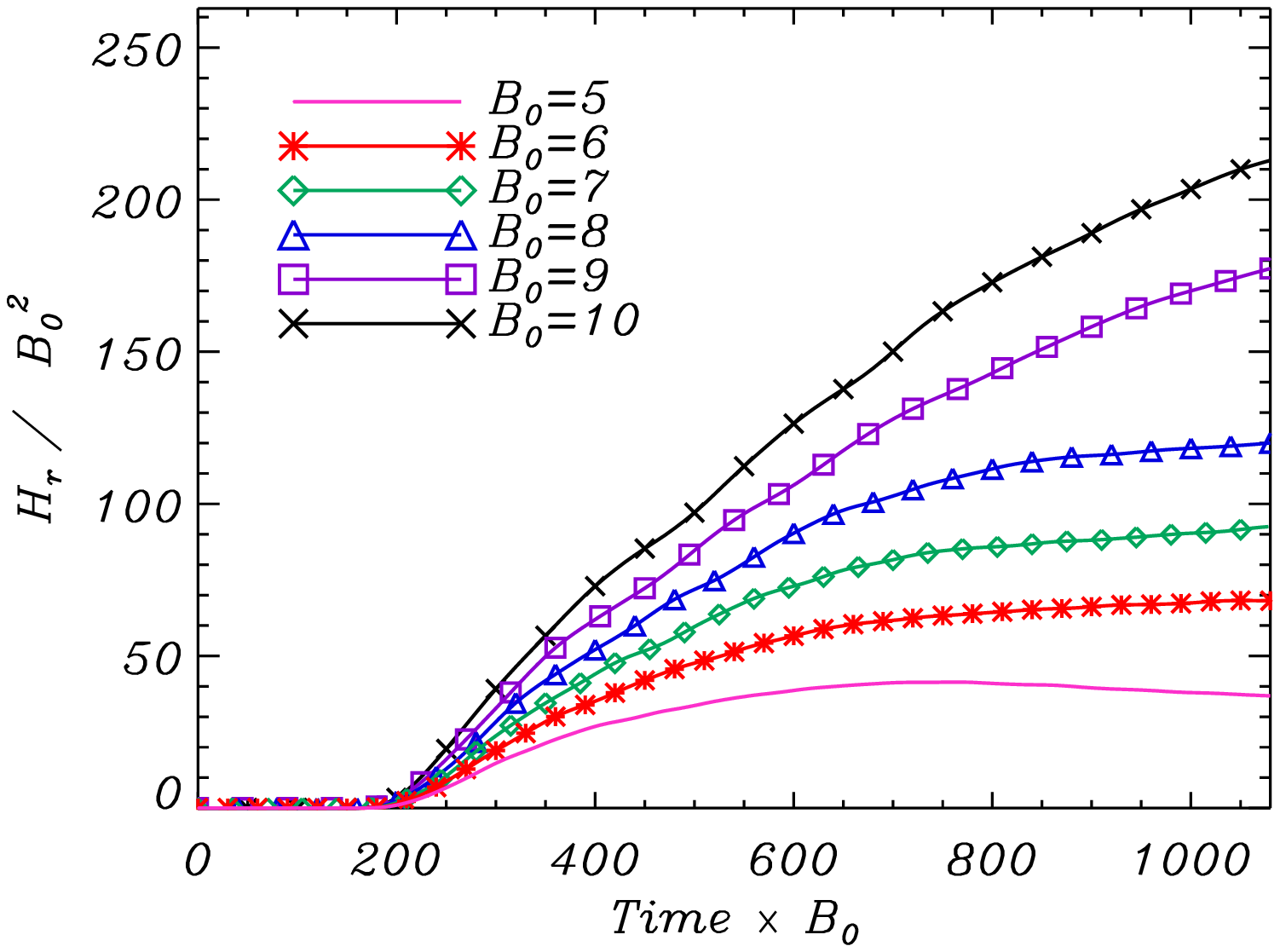}
\subcaption{}
\label{subfig:atmoshelicityb0scaled}
\end{subfigure}
\caption{Relative magnetic helicity calculated within the atmospheric region $z>0$ for various $B_0$ cases. (a) shows the unscaled helicity measured over time and (b) details the rescaled magnetic helicity, $\bar{H_r}=H_r/B_0^2$, measured over rescaled time, $\bar{t}$.}
\label{fig:atmoshelicityb0}
\end{figure}

\begin{table}[ht]
\centering
\caption{Final atmospheric magnetic helicity for Group $1$ experiments.}
\label{table:helicityb0}
\begin{tabular}{cccc}
\hline\hline
\hspace{0.5cm} $B_0$  \hspace{0.5cm}  & \hspace{0.5cm} $H_r$ \hspace{0.5cm} & \hspace{0.5cm} $H_r$ (Wb$^2$) \hspace{0.5cm}  \\ [0.5ex]
 \hline\hline
 $5$ & $922$ & $1.3\times10^{22}$~Wb$^2$ \\[0.5ex]
 $6$ & $2453$ & $3.4\times10^{22}$~Wb$^2$\\[0.5ex]
 $7$ & $4533$ & $6.3\times10^{22}$~Wb$^2$\\[0.5ex]
 $8$ & $7683$ & $1.1\times10^{23}$~Wb$^2$\\[0.5ex]
 $9$ & $14367$ & $2.0\times10^{23}$~Wb$^2$\\[0.5ex]
 $10$ & $21291$ & $5.0\times10^{23}$~Wb$^2$\\[0.5ex]
 \hline
\end{tabular}
\end{table}

From previous work, we expect a linear increase in magnetic helicity in the atmosphere accompanied by a depletion of magnetic helicity in the interior region. This is a result of the direct emergence of flux and rotations of sunspots at the photosphere that twist and stress the atmospheric field while untwisting the interior portion of the field. Figure~\ref{subfig:atmoshelicityb0} considers the temporal evolution of atmospheric helicity for the different cases, and as expected, it increases for all cases. The injection of helicity is clearly ordered by the value of the initial field strength. Both $\mathbf{B}$ and $\mathbf{B_p}$ are directly proportional to $B_0$ in the initial set-up and the vector potentials are therefore also proportional to $B_0$ as evident by the expressions quoted in Paper I. Therefore, the helicity is initially proportional to $B_0^2$. The rescaled atmospheric helicity is plotted against rescaled time in Figure~\ref{subfig:atmoshelicityb0scaled}. There still seems to be a larger amount of $B_0$ scaled helicity for larger $B_0$ cases.  By doubling the initial magnetic field strength of the sub-photospheric tube, $B_0$, the helicity transported to the atmosphere is increased by over $23$ times. As this is a volume integrated quantity, stronger fields occupy a larger portion of the volume, which may explain the larger helicity injection. In addition, the period of strong rotation may inject more helicity than expected by the $B_0^2$ scaling.

\begin{figure}[!ht]
\centering
\begin{subfigure}{0.48\textwidth}
\centering
\includegraphics[scale=0.4]{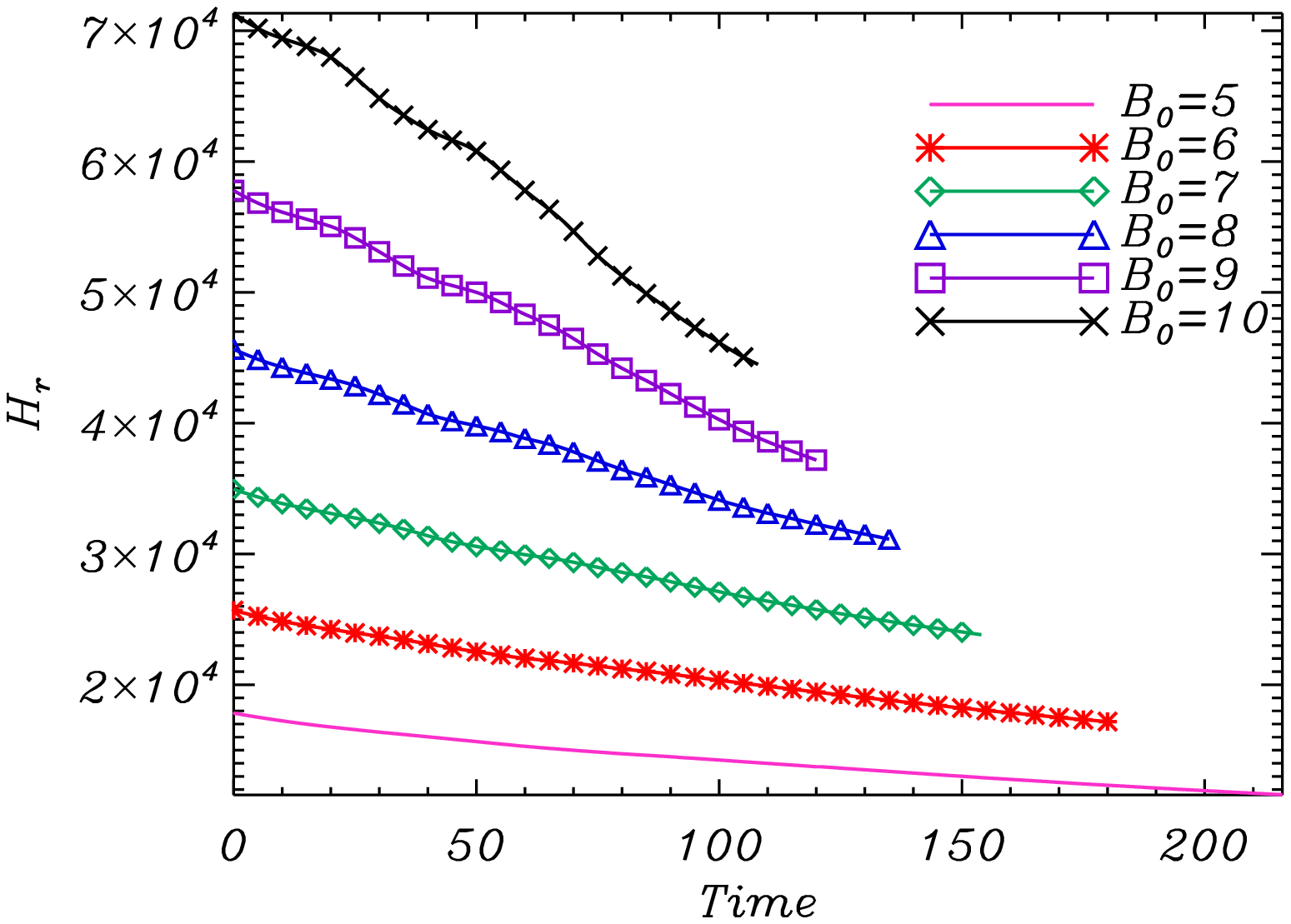}
\subcaption{}
\label{subfig:interiorhelicityb0}
\end{subfigure}\\
\begin{subfigure}{0.48\textwidth}
\centering
\includegraphics[scale=0.4]{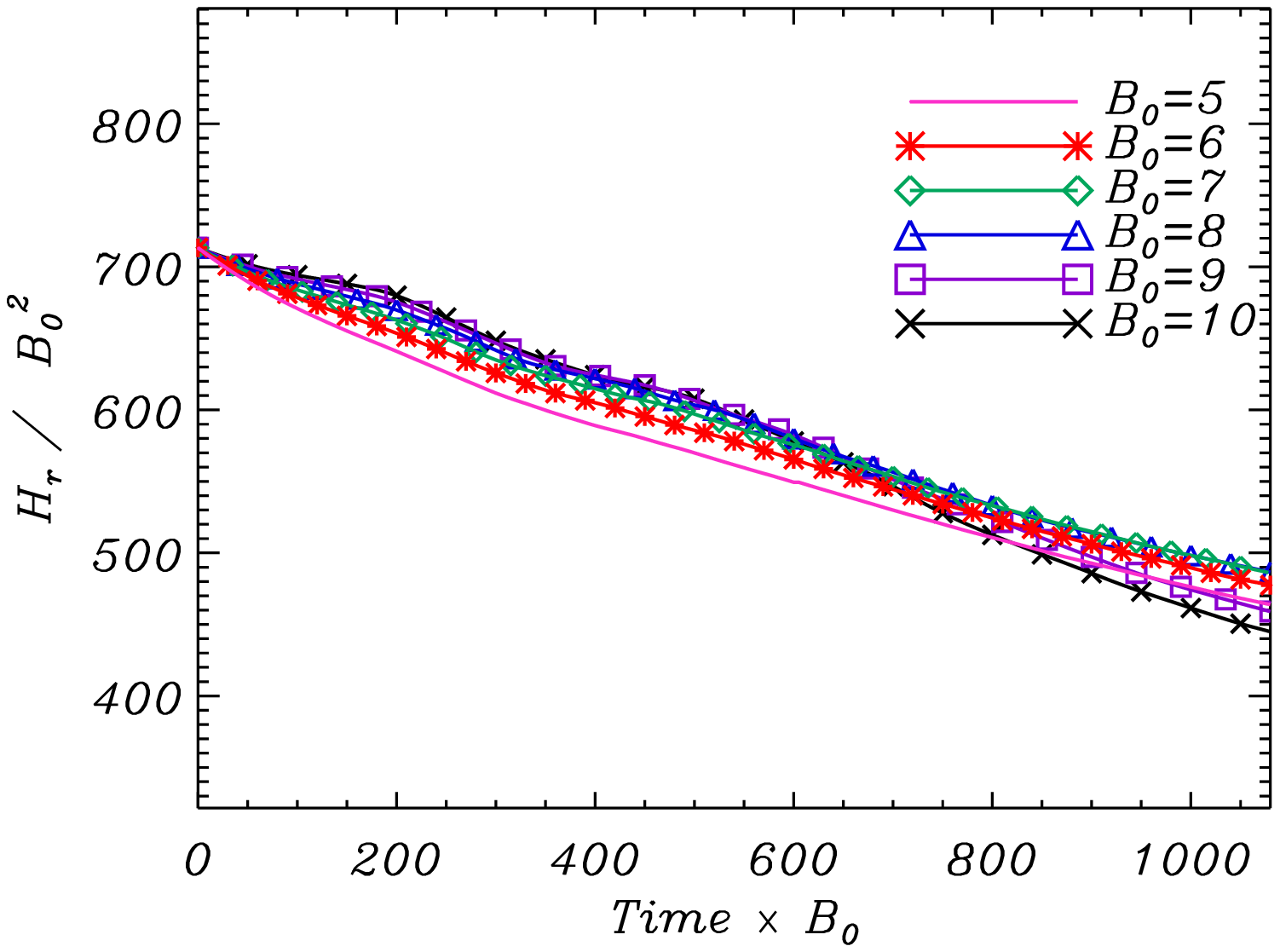}
\subcaption{}
\label{subfig:interiorhelicityb0scaled}
\end{subfigure}
\caption{Relative magnetic helicity calculated within the interior portion $z<0$ for varying $B_0$ cases. (a) considers the unscaled helicity and (b) the rescaled magnetic helicity, $\bar{H_r}=H_r/B_0^2$, measured over rescaled time, $\bar{t}$. }
\label{fig:interiorhelicityb0}
\end{figure}

Similarly, the interior helicity is plotted in Figure~\ref{subfig:interiorhelicityb0}, depicting a reduction in helicity in a similar way. The initial interior helictites are clearly ordered by the value of $B_0$ and can be directly scaled by $B_0^2$. We must bear this in mind when interpreting the helicity as magnetic fields with the same amount of twist may have different amounts of helicity as scaled by their initial magnetic field strength. Furthermore, the stronger experiments clearly undergo a sharper drop-off in helicity. Despite this drop-off, the stronger $B_0$ experiments have considerably larger amounts of interior helicity compared with the weaker $B_0$ experiments during the latter stages of the experiment. For instance, in the $B_0=10$ experiment, there is $6.23\times10^{23}$~Wb$^2$ of helicity left in the interior. This may seem surprising as helicity is a measure of the twist of fieldlines and the fieldline twist is considerably lower for the $B_0=10$ case by the end of the experiment (see Figure~\ref{fig:twistb0}). However, it is important to note that although a non-zero fieldline twist implies a non-zero helicity, the opposite is not necessarily true~\citep{MacTaggart2015}. A magnetic field may have a non-zero helicity without containing a large amount of twist. For example, an untwisted magnetic field may not be exactly potential, and as such may process a non-zero relative helicity.

We again rescale the interior helicity by $B_0^2$ as we have plotted in Figure~\ref{subfig:interiorhelicityb0scaled}. In this case, the helicity seems to demonstrate self-similar behaviour, unlike the atmospheric helicity. The change in helicity by resistive dissipation is much larger in the solar interior, so the scaling with $B_0^2$ shows a better fit. For more details on the dominant contributors to the change in magnetic helicity, including the resistive dissipation term, see~\cite{Sturrock2015} or~\cite{Pariat2015}.

\subsection{Magnetic energy}
\begin{figure}[!ht]
\centering
\begin{subfigure}{0.48\textwidth}
\centering
\includegraphics[scale=0.4]{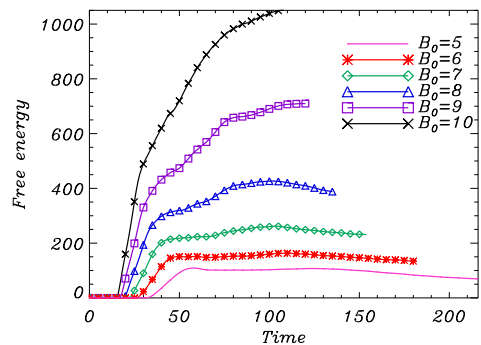}
\subcaption{}
\label{subfig:freeenergyb0}
\end{subfigure}\\
\begin{subfigure}{0.48\textwidth}
\centering
\includegraphics[scale=0.4]{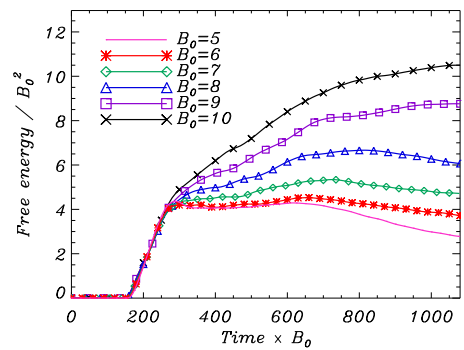}
\subcaption{}
\label{subfig:freeenergyb0scaled}
\end{subfigure}
\caption{Free magnetic energy stored in the atmosphere for Group 1 with (a) the unscaled energy measured over time and (b) the rescaled magnetic energy, $\bar{E_{\text{free}}}=E_{\text{free}}/B_0^2$ measured over rescaled time, $\bar{t}=B_0t$.}
\label{fig:freeenergyb0}
\end{figure}

Finally, an analysis of the magnetic energy is presented to assess how much free energy is produced by the rotational motions at the photosphere. To understand the distribution of energy across the domain, we calculate the free magnetic energy above $z=0$, as follows,
\begin{equation*}
E_{\text{free}} = \int{\frac{\mathbf{B}^2}{2}~\mathrm{d}V} - \int{\frac{\mathbf{B}_p^2}{2}~\mathrm{d}V},
\end{equation*}
where $V$ is defined as the volume above $z=0$. This is essentially the excess energy stored in the field as we have subtracted off the energy stored in the potential field with the same normal distribution on the photospheric boundary. Both the unscaled and scaled free magnetic energy are plotted in Figures~\ref{subfig:freeenergyb0} and~\ref{subfig:freeenergyb0scaled} respectively.
The self-similar evolution is followed strictly for the first $275$ scaled time units. However, later the different $B_0$ cases deviate from the original trend suggesting a different $B_0$ dependence. This agrees with the trend we see in the helicity with larger than expected amounts of energy transported to the atmosphere for stronger initial fields. As the diffusion time is independent of $B_0$, we see greater diffusion for weaker experiments as they have run for a longer unscaled time. This may explain the drop in free magnetic energy in the latter stages of the experiment. The free magnetic energy transported to the atmosphere ranges from $7.2\times10^{21}$~J to $6.9\times10^{22}$~J over the range of $B_0$ values.

\section{Varying $\alpha$ with fixed $B_0$}
\label{sec:varyalpha}
In Group $2$, we fix $B_0$ at $7$ (an axial field strength of $9100$~G) and vary the initial twist of the tube from $\alpha=0.2$ (one turn in $850$~km) to $\alpha=0.4$ (one turn in $425$~km). Using a similar approach to the last group, we pinpoint the effect that the initial twist, $\alpha$, has on the rotation of sunspots. We again investigate a variety of different features related to the rotational movements at the photosphere.

When we modify the degree of twist, $\alpha$, the initial magnetic tension force acting on the tube is affected and this in turn changes the magnetic buoyancy profile of the tube. This is discussed earlier in Section~\ref{sec:parameters} where Figure~\ref{subfig:rhodefalpha} displays a comparison of the density excess for different values of $\alpha$. This reveals that as the value of $\alpha$ is reduced the axial region becomes more buoyant but the surrounding plasma is less buoyant due to the smaller field strength here (see Figure~\ref{subfig:bmagalpha}). Therefore, we need to consider that the buoyancy profile is non-linearly altered in this group suggesting that $\alpha$ may have a more complex effect on the dynamics of the experiment. The non-linear dependence of the density deficit on $\alpha$ makes it difficult to rescale the time to remove this effect as we did in the Group $1$ case. For brevity, we have excluded the methods behind calculating the quantities in this section. Unless stated otherwise, all quantities are calculated as introduced in Group $1$.

\subsection{Rotation angle}
\begin{figure}[ht]
\centering
\includegraphics[scale=0.4]{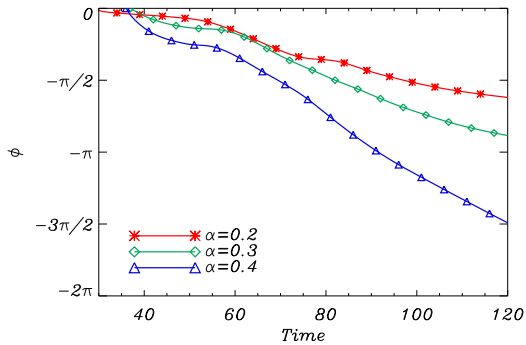}
\caption{Rotation angles measured over time for various $\alpha$ cases, as depicted in the key.}
\label{fig:rotanglesalpha}
\end{figure}

From Figure~\ref{fig:rotanglesalpha}, it is clear that the rotation angle, calculated in the same way as in Section~\ref{sec:rotangleb0}, has some dependence on the initial twist, $\alpha$. Changing the initial degree of twist changes the helical structure of fieldlines and means that fieldlines traced from the same location on the base appear at different locations at the photosphere in all three cases. To try and manage this effect, we have artificially moved all the starting angles to $0$ and the subsequent evolution has been shifted. The time at which the field reaches the photosphere is also affected by $\alpha$ with the $\alpha=0.2$ tube reaching the photosphere first due to its larger axial buoyancy. In Figure~\ref{fig:rotanglesalpha}, there is clearly some trend in that the sunspot in the higher $\alpha$ experiment undergoes a larger rotation. If we surmise that the rotation of sunspots is due to the propagation of twist, and we expect the rotation to attempt to equilibrate the twist imbalance, we predict the rotation angles to be largest for the most highly twisted cases. 

Summarised in Table~\ref{table:alpharesults}, at the end of the section, are the final rotational angles at $t=120$. Notice, in this case all experiments have been run for the same amount of time and we do not rescale $t$. This clearly shows some ordering of the rotation angle with the parameter $\alpha$. By doubling the initial twist, we more than double the final rotation angle. However, the relationship is not as clear as we find in Group $1$ due to the more complicated effect of $\alpha$ on the tube's evolution.

\subsection{Twist}

\begin{figure}[!ht]
 \centering
\includegraphics[scale=0.4]{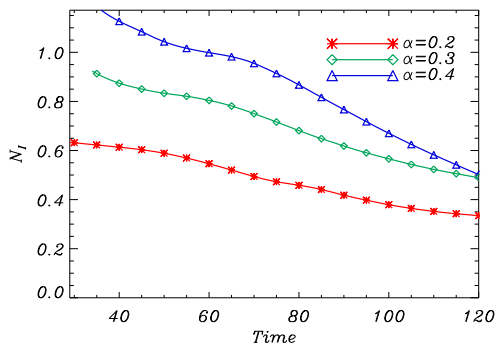}
\caption{Average number of turns, $N_{\text{I}}$, fieldlines undergo within the interior portion of one leg of the tube, measured over time for Group 2 cases.}
\label{fig:twistalpha}
\end{figure}

As discussed in the previous section, the rotational motions at the photosphere extract twist from the interior. The number of turns the field takes around the axis within an interior section of the tube, $N_{\text{I}}$, is presented in Figure~\ref{fig:twistalpha} as calculated in Equation~\eqref{eq:turnsinterior}. Given that all experiments start with a different initial twist, $\alpha$, they all contain differing amounts of twist within the interior when they first intersect the photosphere. In addition, weakly twisted tubes reach the photosphere first as they are more buoyant. Due to the larger rotation seen for more highly twisted fields, the interior twist is extracted more efficiently.

\subsection{Vorticity}
\begin{figure}[!ht]
 \centering
\includegraphics[scale=0.4]{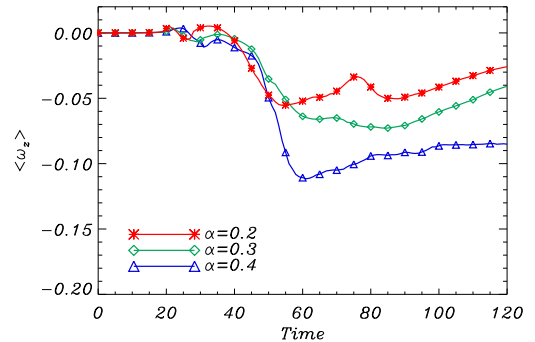}
\caption{Average vorticity over time for various $\alpha$ cases.}
\label{fig:vorticityalpha}
\end{figure}

As the flux tubes reach the photosphere, vortical motions develop on the sunspot centres, as shown in Figure~\ref{fig:vorticityalpha}. As $B_0$ is constant throughout this group of simulations, the region where $B_z > 3/4\text{max}(B_z)$ is approximately constant due to $B_z$'s weak dependence on $\alpha$. There is, however, a clear trend indicating that tubes with a higher initial degree of twist have larger vortical motions developing on their sunspot centres. This is expected as we predict vortical motions at the photosphere untwist interior field in an attempt to equilibrate the twisted interior with the stretched atmospheric field. If the initial flux tube is highly twisted, the fieldlines threading through the sunspot must rotate through a larger angle (see Figure~\ref{fig:rotanglesalpha} and Table~\ref{table:alpharesults}), producing a higher magnitude of vorticity.

\subsection{Magnetic helicity}

\begin{figure}[!ht]
\centering
\includegraphics[scale=0.4]{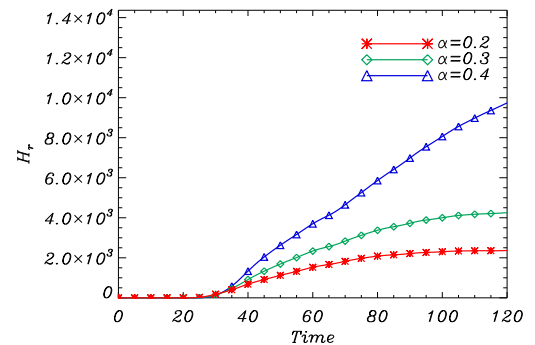}
\caption{Relative magnetic helicity calculated above $z=0$ in the atmospheric portion of the volume for varying $\alpha$ cases.}
\label{fig:atmoshelicityalpha}
\end{figure}

\begin{figure}[!ht]
\centering
\includegraphics[scale=0.4]{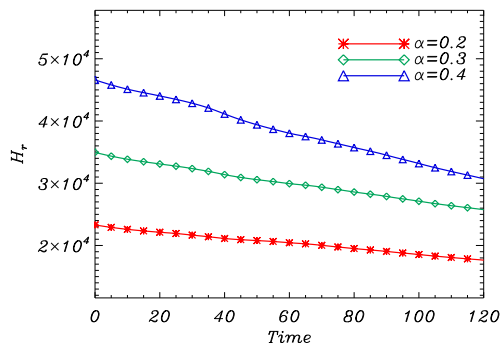}
\caption{Relative magnetic helicity calculated within the interior portion below $z=0$ for varying $\alpha$ cases.}
\label{fig:interiorhelicityalpha}
\end{figure}

To complete our analysis of the Group 2 simulations, we present an analysis of the magnetic helicity within two distinct sections of the domain, namely the solar interior and atmosphere, as separated by the $z=0$ photospheric boundary. The temporal evolution of this quantity, as calculated using Equation~\eqref{eq:helicity}, is shown in Figures~\ref{fig:atmoshelicityalpha} and~\ref{fig:interiorhelicityalpha} for the atmosphere and interior respectively.
As expected, there is a linear increase in the atmospheric helicity in all cases due to the emergence of flux into the atmosphere and the twisting of the atmospheric field caused by photospheric horizontal flows. The degree of twist clearly alters the amount of magnetic helicity transported to the atmosphere. The amount of helicity in the atmosphere tends to saturate much more quickly for the lower twist cases (red and green). The highly twisted (blue) case, on the other hand, continues to increase over the whole experiment. The final helicity values for the three cases are summarised in Table~\ref{table:alpharesults} in physical units. There is some evidence that the helicity may be proportional to $\alpha^2$. However, there is not a direct relationship given the non-linear dependence of the magnetic field on $\alpha$, made clear by the initial field, $\mathbf{B}$, outlined in Equation~\eqref{eq:initialb}.

The initial magnetic helicity stored in the interior is clearly altered by the degree of initial twist of the field as evidenced in Figure~\ref{fig:interiorhelicityalpha}. Initially, the total magnetic helicity within the volume is given solely by the magnetic helicity stored within the interior as the magnetic flux tube is yet to emerge and enter the atmosphere. The starting interior helicities range from $3.25\times10^{23}$~Wb$^2$ for the $\alpha=0.2$ case to $6.5\times10^{23}$~Wb$^2$ for the $\alpha=0.4$ case. 
The rate at which the helicity in the interior decreases also appears to be dependent on the initial twist with a steeper decline for the highest $\alpha$ experiment. This directly links to the more rapid rotation observed in the highly twisted case. 

\begin{table}[ht]
\centering
\caption{Summary of key results at $t=120$ for Group 2 experiments.}
\label{table:alpharesults}
\begin{tabular}{cccc}
\hline\hline
\hspace{0.4cm} $\alpha$ \hspace{0.4cm} & \hspace{0.4cm} $\phi$ \hspace{0.4cm}  & \hspace{0.4cm} $H_r$ (Wb$^2$) \hspace{0.4cm} & \hspace{0.4cm} $E_{\text{free}}$ (J) \hspace{0.4cm}  \\ [0.5ex]
 \hline\hline 
 $0.2$ & $112\degree$ & $3.3\times10^{22}$~Wb$^2$ & $1.1\times10^{22}$~J   \\[0.5ex]
 $0.3$ & $177\degree$ & $6.0\times10^{22}$~Wb$^2$ & $1.6\times10^{22}$~J\\[0.5ex]
 $0.4$ & $268\degree$ & $1.4\times10^{23}$~Wb$^2$ & $3.4\times10^{22}$~J\\[0.5ex]
 \hline
\end{tabular}
\end{table}

\section{Conclusions}
\label{sec:conclude}

In this paper, we presented results from $3$D MHD simulations of buoyant twisted toroidal flux tubes as they rise through the solar interior and emerge into the atmosphere. Our primary aim was to investigate the rotation of the photospheric footpoints. We varied the magnitudes of two parameters governing the magnetic structure of the tube, namely the axial magnetic field strength, $B_0$, and the twist, $\alpha$, of the sub-photospheric flux tube. Our focus was to identify the distinct effect of each of these parameters on the rotational motion at the photosphere and the many ramifications of this photospheric velocity.

To investigate this effect, we analysed various quantities relating to the plasma and magnetic field. To directly measure the rotation, we calculated the rotation angle based on the axis of the flux tube as the centre of the sunspot. This allowed us to make a direct comparison of how magnetic field strength and twist affect rotation rates. Similarly, we looked at how the field strength and twist affect the plasma vorticity within the sunspots. In addition, we analysed the twist of individual fieldlines, magnetic energy, and helicity to study the twist and writhe contained within the different sub-volumes of the domain. This allowed us to understand the distribution of twist across the system, and the transport of twist from the interior to atmosphere of the model.

Many interesting relationships were found for Group $1$ in which we kept the twist, $\alpha$, constant and varied the axial magnetic field strength, $B_0$. This parameter investigation provides us with an insight into how the initial magnetic field strength affects the amount by which the flux tube rotates. Surprisingly, we found the vertical photospheric magnetic field strength to scale with $B_0^{2}$ when we varied the initial axial field strength, $B_0$, of the sub-photospheric field. All components of the magnetic field were initially proportional to $B_0$, but by the time the tube reached the photosphere the magnetic field's magnitude and direction were adapted as governed by the initial $B_0$. Stronger fields tended to emerge more fully with a vertically directed axis, whereas weaker fields tended to spread horizontally at the photosphere to allow the magnetic buoyancy instability to initiate. In addition, stronger fields extended higher into the atmosphere and possessed a greater axial length. In brief, the magnetic field was altered on its journey to the photosphere in all experiments, and hence the results we found may be surprising and unforeseen.

Another particularly interesting result we found in Group 1 was that the rotation angle is dependent on $B_0$. The timescale over which the rotation occurred is dependent on $B_0$ owing to the density deficit's dependence on $B_0$. Hence, in a fixed time, a larger rotation angle was passed through by sunspots in higher $B_0$ experiments. To remove this time dependence, we scaled the time as $\bar{t}=t B_0$, but this did not reveal a self-similarity. Instead, we discovered that the rescaled rotation angle, $\bar{\phi}=\phi / B_0$, is self-similar. This result is surprising as we may have expected the final rotation angle to be the same for varying magnetic field strength as the initial fields shared the same twist and helical structure. The basis for this relationship was difficult to ascertain from this result alone, but became clearer when we investigated the twist. It is conceivable that if we performed the experiments for a longer time, the rotation would cease for stronger experiments and continue for weaker experiments until a plateau was reached for all cases. Owing to the diffusion timescale and computational expense, we are currently unable to check this. However, this seems unlikely as the rotation rate dropped off significantly to almost zero by the end of the weaker $B_0$ experiments. If the rotation rate does cease for the weaker field experiments and there is not any latter difference in rotation angle for later times, it is possible that the magnetic fields in the weaker experiments are unable to extract as much interior twist as the stronger experiments.

The investigations of twist were also in agreement with the rotation angle results. By considering the fieldline twist and the force-free parameter $\alpha_L$, we found a considerable amount of twist left in the interior for the weaker experiments. Based on an idealised analytic model,~\cite{Longcope2000} suggested that the photospheric footpoints of an emerging tube will rotate until the twist per unit length equilibrates along the length of the field, with the interior $\alpha_L$ matching the coronal $\alpha_L$. In our experiments we do see some evidence of $\alpha_L$ heading to a constant, but found a higher interior $\alpha_L$ for weaker experiments and a lower interior $\alpha_L$ for stronger experiments compared to the constant coronal value. In order to test Longcope's theory we performed a separate idealised experiment (not presented in this paper), composed of a twisted sub-photospheric flux tube connected to a straight coronal field in a simple stratified domain, and allowed it to evolve. We found that a torsional Alfv\'{e}n wave propagated twist from the interior to the corona, and importantly found that this process continued until the twist per unit length equilibrated. Interestingly, we found that the photospheric footpoints underwent an over-rotation as the rate of twist in the interior dropped below that of the coronal value. This simple experiment helps us to understand how the flux tubes in our parameter study would have behaved if we had been able to run the experiments for a longer time. Based on this, we expect the sunspots in stronger experiments to rotate in the opposite sense and to return to a constant twist rate, and expect the sunspots in weaker experiment to continue to rotate. However, as the axis fieldline is shorter for weaker fields, the final twist per unit length is larger so it is conceivable that even once $\alpha_L$ has equilibrated, there may still be a considerable amount of twist left in the interior. This could help us explain why the rotation angles are smaller for weaker experiments. In addition, we found the helicity and magnetic energy to be ordered by $B_0$. In stronger field experiments, we noticed a larger transport of magnetic energy and helicity to the atmosphere.

Varying the initial degree of twist also had an effect on the amount of rotation we calculated within the sunspots. However, it was difficult to find direct relationships with $\alpha$, given the non-linear dependence of the initial field on $\alpha$ and that the magnetic buoyancy profile is also altered by the degree of twist. We found the rotation angle, twist, helicity, and vorticity to be ordered by the degree of twist, $\alpha$. Larger vortical motions developed in the highly twisted experiment, transporting more helicity into the atmosphere. Although not included in this paper, we also found a larger increase in free magnetic energy in the atmosphere for the highly twisted field as depicted in Table~\ref{table:alpharesults}. Work must be done to understand the non-linear effect of the twist on the evolution of the tube. However as the twist and magnetic tension force are inherently linked non-linearly it is difficult to scale quantities in a simple linear manner.

As mentioned in Paper I, it should be noted that we analysed particularly small active regions in our experiments and, hence, the rotation angles we observed occur over much shorter timescales than those found in observations. If we scaled up our experiments, we expect the timescales of rotation, and hence the rotation rate, to be comparable with observations. However, as the size of active region was kept constant across all experiments, this did not impact on this parameter study. 

Varying the magnetic field strength and twist of the interior flux tube has a profound effect on the evolution of the flux tube in our experiments as well as the rotational properties at the photosphere. Although this paper has provided insight into this effect, further questions have been raised. Are the trends we find in our numerical parameter investigation seen in observations? For instance, are weaker fields unable to release the twist stored within the interior? If so, what are the reasons behind this? We predict that sunspots stop rotating once the interior $\alpha_L$ balances the coronal $\alpha_L$~\citep{Longcope2000} and believe it is the height of the axis and in turn length of the fieldlines that hinder the transport of twist in weaker magnetic fields. We would like to run experiments for longer in order to study the final rotation angle and distribution of twist across the system. There is much work to be done here from both an observational and modelling point of view. Furthermore, we assumed that the atmosphere is unmagnetised in our experiments but this is certainly not the case within the Sun's corona. It would be interesting to investigate whether the addition of a magnetised corona affects the rate and amount of rotation within these experiments. The insights found in this paper should be tested and compared with future observations as a model to predict how both the initial magnetic field strength and twist of a sub-photospheric flux tube affect the level of rotation within sunspots. 

\begin{acknowledgements}
ZS acknowledges the financial support of the Carnegie Trust for Scotland. This work used the DIRAC 1, UKMHD Consortium machine at the University of St Andrews  and the DiRAC Data Centric system at Durham University, operated by the Institute for Computational Cosmology on behalf of the STFC DiRAC HPC Facility (\href{http://www.dirac.ac.uk}{www.dirac.ac.uk}). This equipment was funded by BIS National E-infrastructure capital grant ST/K00042X/1, STFC capital grant ST/H008519/1, and STFC DiRAC Operations grant ST/K003267/1 and Durham University. DiRAC is part of the National E-Infrastructure. 
\end{acknowledgements}
\bibliographystyle{aa}
\bibliography{MyCollection}

\end{document}